\documentclass[10pt,final]{IEEEtran}

\usepackage{amsmath, amsfonts, amssymb}
\usepackage{bbm}
\usepackage{bbold}
\usepackage{graphicx,dblfloatfix}
\usepackage[space]{cite}
\usepackage{hyperref}
\usepackage{color}
 
\newtheorem{theorem}{Theorem}

\newtheorem{lemma}[theorem]{Lemma}

\newtheorem{definition}{Definition}

\newtheorem{remark}{Remark}

\newcommand{\set}[1]{\mathcal{#1}}
\renewcommand{\b}[1]{\mathbb{#1}}
\newcommand{\step}[2]{\stackrel{\textnormal{#1}}{#2}}
\newcommand{\indicator}[1]{\mathbbm{1}{\left\{ {#1} \right\} }}

\renewcommand{\Pr}{\mathbb{P}}
\newcommand{\E}{\mathbb{E}}
\newcommand{\es}{{\epsilon}}
\newcommand{\ec}{{\delta}}
\newcommand{\eh}{{\epsilon_\text{h}}}
\newcommand{\R}{\boldsymbol{R}}
\newcommand{\phib}{\boldsymbol{\phi}}

\newcommand{\D}{\boldsymbol{D}}
\newcommand{\ns}{{n_\text{s}}}
\newcommand{\nc}{{n_\text{c}}}
\newcommand{\nh}{{n_\text{h}}}

\newcommand{\rt}[1]{{\color{black}{#1}}}

\newcommand{\maw}[1]{{\color{black}{#1}}}

\allowdisplaybreaks[4] 
\graphicspath{{.}{./Figures/}} 

\begin{document}

\title{\rt{Slepian-Wolf Coding for Broadcasting with\\ Cooperative Base-Stations}}

\author{Roy Timo and Mich\`{e}le Wigger
\thanks{R.~Timo is an Alexander von Humboldt research fellow with the Technische Universit\"{a}t M\"{u}nchen, e-mail roy.timo@tum.de.}
\thanks{M.~Wigger is with Telecom ParisTech, e-mail michele.wigger@telecom-paristech.fr.}
\thanks{This work was supported by the city of Paris under the programme ``Emergences'' and the Alexander von Humboldt Foundation.}
}

\maketitle

\begin{abstract}
We propose a base-station (BS) cooperation model for broadcasting a \rt{discrete memoryless source} in a cellular or heterogeneous  network. The model allows the receivers to use helper BSs to improve network performance, and it permits the receivers to have prior side information about the \rt{source}. \rt{We establish the model's information-theoretic limits in two operational modes: In Mode 1, the helper BSs are given information about the channel codeword transmitted by the main BS, and in Mode 2 they are provided correlated side information about the source. Optimal codes for Mode 1 use \emph{hash-and-forward coding} at the helper BSs; while, in Mode 2, optimal codes use source codes from Wyner's \emph{helper source-coding problem} at the helper BSs. We prove the optimality of both approaches by way of a new list-decoding generalisation of~\cite[Thm.~6]{Tuncel-Apr-2006-A}, and, in doing so, show an operational duality between Modes 1 and 2.}
\end{abstract}


\section{Introduction \& Main Reults}\label{Sec:S&MR}

\IEEEPARstart{T}{he} proliferation of wireless communications devices presents significant performance challenges for cellular networks, and it will require more sophisticated heterogeneous networks in the near future~\cite{Damnjanovic-Jun-2011-A,Ghosh-Jun-2012-A}. A powerful methodology for improving performance is centered on the idea of base-station (BS) cooperation: Instead of operating independently, future BSs will coordinate encoding and decoding operations using information shared over backbone networks. \rt{The tremendous potential of BS cooperation has been widely investigated~\cite{Karakayali-Aug-2006-A,Gesbert-Dec-2010-A,Andrews-Apr-2005-A};} however, despite many advances, there remains significant challenges in understanding and exhausting the benefits of cooperation. Indeed, the fundamental limits of cooperation are fully understood in very few settings~\cite{Gesbert-Dec-2010-A}. 

\begin{figure}
\begin{center}
\includegraphics[width=0.4\textwidth]{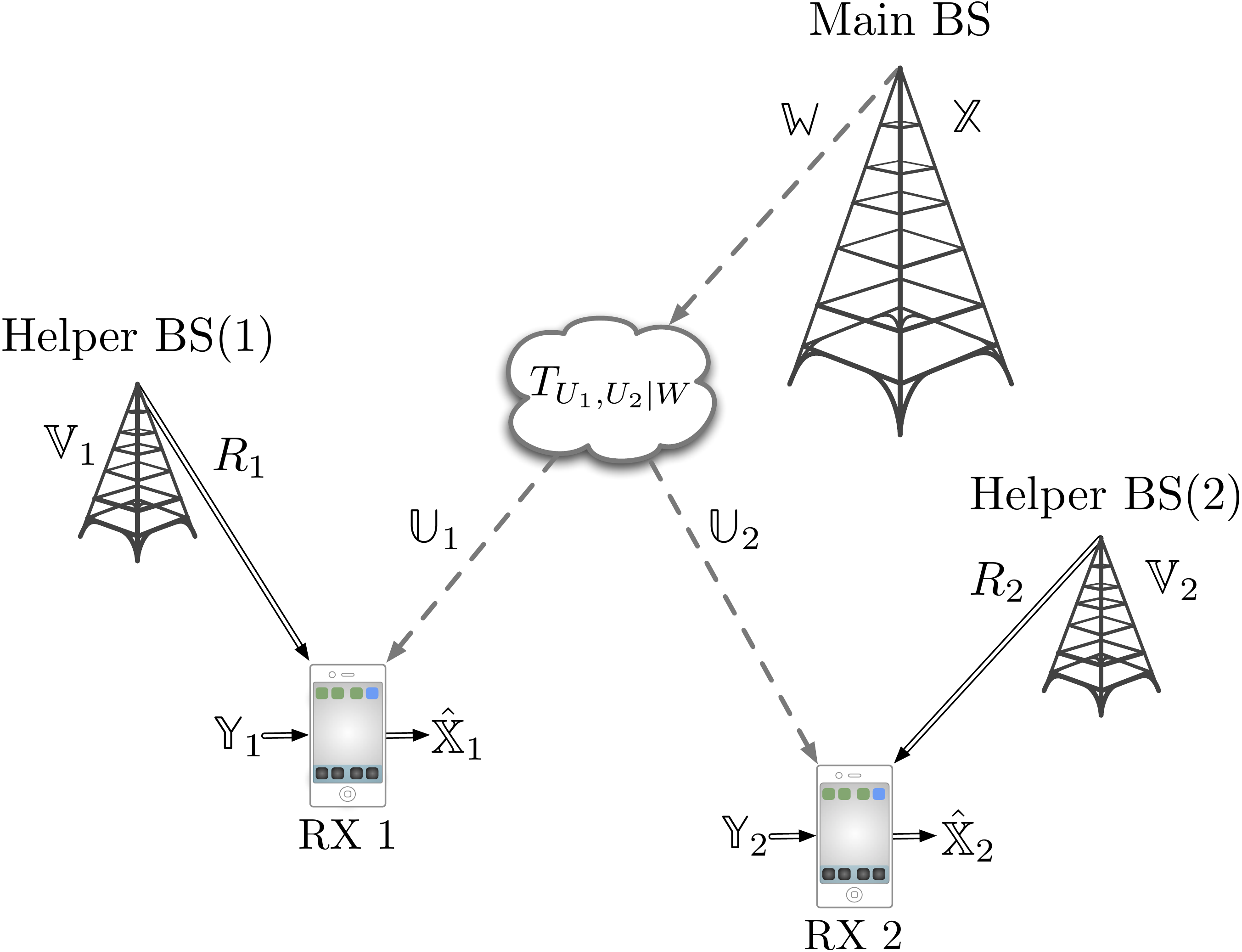}
\caption{Broadcasting with helper BSs and receiver side information.}
\label{Fig:HSW-Broadcast}
\end{center}
\end{figure}

To help understand the full potential of BS cooperation, we consider a simple, but rather useful, broadcast model. The setup for two receivers is shown in Figure~\ref{Fig:HSW-Broadcast}. A \rt{source} $\b{X}$ is to be reliably transmitted over a broadcast channel to many receivers, and the idea is to improve network performance by allowing the receivers to be assisted by \emph{helper} BSs. In a future heterogenous network, for example, the helpers may be pico or femto BSs operating within the main macro cell on orthogonal channels~\cite{Chandrasekhar-Oct-2009-A}. Alternatively, the helpers may be WiFi hotspots through which traffic is diverted from a heavily loaded cellular network~\cite{Singh-2014-A}. The purpose of this paper is to characterise the model's information-theoretic limits, and to provide architectural insights for optimal codes. 

We assume that the broadcast channel from the main BS is discrete and memoryless, and the channels from the helper BSs are noiseless and rate-limited. Although this setup does not capture all modes of cooperation, it nevertheless has enough sophistication to provide insight into some important coding challenges.  For example, consider the idea of augmenting traffic flow in a cellular network via a WiFi hotspot: The hotspot's radio-access technology is orthogonal to that of the cellular network, and a cellular network engineer can well approximate the WiFi link by a noiseless rate-limited channel. A natural question is then: What coding techniques at the BSs and WiFi hotspot yield the best overall performance?

Within the above framework, we consider two operational modes.
\begin{itemize}
\item \emph{Mode~1:} The helper BSs are given side information about the channel codeword transmitted by the main BS. 
\item \emph{Mode~2:} The helper BSs are given correlated side information about the \rt{source} $\b{X}$.
\end{itemize}
We will see that optimal codes for Mode~1 combine virtual-binning from \emph{Slepian-Wolf Coding over Broadcast Channels}~\cite{Tuncel-Apr-2006-A} with hash-and-forward coding for the \emph{primitive relay channel}~\cite{Kim-Mar-2008-A}. Optimal codes for Mode~2, on the other hand, combine virtual binning with source codes from \emph{Wyner's helper side-information problem}~\cite{Wyner-May-1975-A}. We prove the optimality of both codes by way of a new list-decoding generalisation of~\cite[Thm.~6]{Tuncel-Apr-2006-A}, and, in doing so, show an operational duality between Modes 1 and 2. 

The paper is organised as follows. The BS cooperate model is defined in Section~\ref{Sec:Prelim}, and our results are  summarised in Section~\ref{Sec:LimitsCoop}. We introduce and solve a list-decoding broadcast problem in Sections~\ref{Sec:LD} through~\ref{Sec:Proof:Lem:SW-List:Ach}. Finally, we prove the BS cooperation results in Sections~\ref{Sec:Proof:Thm:M1:Con} to~\ref{Sec:Proof:Thm:M2:Ach}.


\section{Preliminaries}\label{Sec:Prelim}


\subsection{Notation}

We denote random variables by uppercase letters, e.g. $A$; their alphabets by calligraphic typeface, e.g. $\set{A}$; and elements of an alphabet by lowercase letters, e.g. $a \in \set{A}$. The cartesian product of alphabets $\set{A}$ and $\set{B}$ is $\set{A} \times \set{B}$, and the $n$-fold cartesian product of $\set{A}$ is $\set{A}^n$. When $n$ is clear from context, we use boldface notation for a sequence of $n$ random variables on a common alphabet, e.g. $\b{A} = (A_1,A_2,\ldots,A_n) \in \set{A}^n$.


\subsection{Source and Channel Setup}

The main BS is required to communicate a \rt{source}
\begin{equation*}
\b{X} = (X_1,X_2,\ldots,\rt{X_\ns})
\end{equation*}
over a discrete memoryless broadcast channel to $K$ receivers with side information; the side information at receiver $k$, for $k \in \{1,2,\ldots,K\}$, is denoted by
\begin{equation*}
\b{Y}_k = (Y_{k,1},Y_{k,2},\ldots,\rt{Y_{k,\ns}}).
\end{equation*}
For example, $\b{X}$ and $\b{Y}_k$ may be the current and previous states of a mobile application, the global and local contents of a cloud storage drive, or the current and previous frames of a video feed. Alternatively, specific choices of $\b{X}$ and $\b{Y}_k$ lead to the bi-directional broadcast channel and complementary side information model~\cite{Oechtering-Jan-2008-A,Kramer-Sep-2007-C,Wyner-Jun-2002-A,Timo-Nov-2010-A}. For generality, let us only assume that the \rt{source} and side information are emitted by a discrete memoryless source\footnote{It is possible to extend this research to discrete ergodic sources using, for example, the methods of~\cite{Cover-Mar-1975-A}. However, discrete memoryless sources lead to more instructive proofs with less technical and notational difficulties.}. That is,  
\begin{equation*}
(\b{X},\b{Y}_1,\b{Y}_2,\ldots,\b{Y}_K) := \big\{(X_i,Y_{1,i},Y_{2,i},\ldots,Y_{K,i}) \big\}_{i = 1}^{\rt{\ns}}
\end{equation*}
is a sequence of $\rt{\ns}$ independent and identically distributed (iid) source/side-information tuples $(X,Y_1,Y_2,\ldots,Y_K)$ defined by a fixed, but arbitrary, joint probability mass function (pmf) on the Cartesian product space $\set{X} \times \set{Y}_1 \times \set{Y}_2 \times \cdots \times \set{Y}_K$.

Let $\set{W}$ denote the broadcast channel's input alphabet and $\set{U}_k$ its output alphabet at receiver~$k$. The main BS transmits
\begin{equation*}
\b{W} := f(\b{X})
\end{equation*}
over the broadcast channel, where $f : \rt{\set{X}^\ns} \longrightarrow \rt{\set{W}^\nc}$ is the BS's encoder and $\b{W} = (W_1,W_2,\ldots,\rt{W_\nc})$ is a codeword with \rt{$\nc$} symbols. \rt{The ratio of channel symbols to source symbols, 
\begin{equation*}
\kappa := \frac{\nc}{\ns},
\end{equation*}
is called the \emph{bandwidth expansion factor}.}

Receiver $k$ observes $\b{U}_k = (U_{k,1},U_{k,2},\ldots,\rt{U_{k,\nc}})$ from the channel. The channel outputs, across all receivers, conditionally depend on the codeword $\b{W}$ via the memoryless law 
\begin{multline*}
\Pr[\b{U}_1=\b{u}_1,\b{U}_2=\b{u}_2,\ldots,\b{U}_K=\b{u}_K|\b{W}=\b{w}]\\
= \prod_{i=1}^{\rt{\nc}} T(u_{1,i},u_{2,i},\ldots,u_{K,i}|w_i), 
\end{multline*}
where $\b{w} \in \rt{\set{W}^\nc}$, $\b{u}_k \in \rt{\set{U}_k^{\nc}}$ and $T(u_1,\ldots,u_K|w)$ is a fixed, but arbitrary, conditional probability.


\subsection{No Base-Station Cooperation}

Momentarily suppose that there is no BS cooperation, and that the \rt{source} is to be losslessly reconstructed using only the channel outputs and side information at each receiver. In this setting, reliable communication is possible if (and only if)\footnote{Replace the strict inequality in~\eqref{Eqn:Tuncel} with an inequality.} there exists a pmf $P_W$ on $\set{W}$ such that~\cite{Tuncel-Apr-2006-A} 
\begin{equation}\label{Eqn:Tuncel}
H(X|Y_k) <  \rt{\kappa} I(W;U_k),\quad \forall\ k,
\end{equation}
where $(W,U_1,U_2,\ldots,U_K) \sim P_W(\cdot) T(\cdot|\cdot)$. The necessity and sufficiency of~\eqref{Eqn:Tuncel} for reliable communication is an elegant and powerful result with applications throughout network information theory; for example, consider~\cite{Oechtering-Jan-2008-A, Kramer-Sep-2007-C, Timo-Nov-2010-A, Wyner-Jun-2002-A} and \cite{Gunduz-Sep-2009-A,Nayak-Apr-2010-A1,Timo-Feb-2013-A,Gunduz-Apr-2013-A}. Indeed, a new list-decoding generalisation of~\eqref{Eqn:Tuncel} will play a central role in this paper.


\subsection{Base-Station Cooperation}
Let us now return to the BS cooperation model. The helper BS of receiver $k$, denoted BS($k$), obtains \rt{side} information 
\begin{equation*}
\b{V}_k = (V_{k,1},V_{k,2},\ldots,\rt{V_{k,\nh}})
\end{equation*}
about the \rt{source $\b{X}$} or the codeword \rt{$\b{W}$} via a backbone network. \rt{Here $\nh = \ns$ (resp. $\nh = \nc$) when BS($k$) has side information about $\b{X}$ (resp.~$\b{W}$)}, and a precise definition of $\b{V}_k$ will \rt{be} given shortly. BS($k$) sends  
\begin{equation*}
M_k := f_k(\b{V}_k)
\end{equation*}
over a noiseless channel to receiver $k$, where $f_k: \rt{\set{V}^\nh} \rightarrow \{1,2,$ $\ldots,\lfloor 2^{\rt{\ns} R_k} \rfloor\}$ is BS($k$)'s encoder and $R_k$ is its rate \rt{(in bits per source symbol\footnote{Here we have synchronised the rate $R_k$ to the number of source symbols $\ns$. Alternatively, one could synchronise $R_k$ to the number of channel symbols by replacing $\ns$ with $\nc$ in the definition of $f_k$.})}. Receiver~$k$ attempts to recover the \rt{source} via
\begin{equation*}
\hat{\b{X}}_k := g_k(\b{U}_k,\b{Y}_k,M_k),
\end{equation*}
where $g_k : \rt{\set{U}_k^\nc} \times \rt{\set{Y}_k^\ns} \times \{1,2,\ldots,\lfloor 2^{\rt{\ns} R_k} \rfloor\} \longrightarrow \rt{\set{X}^\ns}$ is the receiver's decoder. The collection of all encoders and decoders is called an $(\rt{\ns,\nc},R_1,R_2,\ldots,R_K)$-\emph{code}. 


\subsection{Mode 1 (helper side information about the codeword $\b{W}$)}

Suppose that $\b{V}_k$ is the entire codeword $\b{W}$ or a scalar quantised version thereof. Quantisation is appropriate, for example, when the backbone network is rate limited. More formally, let $\phi_k : \set{W} \rightarrow \set{V}_k$ be an arbitrary but given deterministic mapping (scalar quantiser) and  
\begin{equation*}
V_{k,i} := \phi_k(W_i),\quad \forall\ i.
\end{equation*}
The main problem of interest is to determine when reliable communication is achievable in the following sense. 

\rt{
\begin{definition}\label{Def:BSCoopM1}
Fix the  bandwidth expansion factor $\kappa$, helper BS rates $\R := (R_1,R_2,\ldots,R_K)$, and scalar quantisers $\phib := (\phi_1,\phi_2,\ldots,\phi_K)$. We say that a source/side information tuple $(X,Y_1,Y_2,$ $\ldots,Y_K)$
is $(\kappa,\R,\phib)$-\emph{achievable} if for any $\epsilon > 0$ there exists an $(\ns,\nc,R_1,R_2,\ldots,R_K)$-code such that  
\begin{equation}\label{Eqn:Def:BSCoop}
\frac{\nc}{\ns} = \kappa
\quad \text{and} \quad
\Pr[\hat{\b{X}}_k \neq \b{X}] \leq \epsilon,\quad \forall\  k,
\end{equation}
holds for sufficiently large $\ns$ and $\nc$. 
\end{definition}
}

\subsection{Mode 2 (helper side information about the source $\b{X}$)}

Suppose that  $\b{V}_k$ is directly correlated with the \rt{source} and side information. That is, assume $(\b{X},\b{Y}_1,\b{Y}_2,\ldots,\b{Y}_K,\b{V}_1,\b{V}_2,$ $\ldots,\b{V}_K)$ is emitted by an arbitrary discrete memoryless source and thus is a sequence of $\rt{\ns}$ iid tuples $(X,Y_1,Y_2,\ldots,Y_K,V_1,$ $V_2,\ldots, V_K$). We are interested in the following definition of achievability.
\rt{
\begin{definition}\label{Def:BSCoopM2}
Fix the  bandwidth expansion factor $\kappa$ and helper BS rates $\R := (R_1,R_2,\ldots,R_K)$. We say that a source/side information tuple $(X,Y_1,Y_2,\ldots,Y_K,V_1,V_2,\ldots,$ $V_K)$
is $(\kappa,\R)$-\emph{achievable} if for any $\epsilon > 0$ there exists an $(\ns,$ $\nc,R_1,R_2,\ldots,R_K)$-code such that~\eqref{Eqn:Def:BSCoop} holds for sufficiently large $\ns$ and $\nc$. 
\end{definition}
}


\section{Information-Theoretic Limits of \\ BS Cooperation}\label{Sec:LimitsCoop}

We now give necessary and sufficient conditions for a source/side-information tuple to be achievable in the sense of Definitions~\ref{Def:BSCoopM1} and~\ref{Def:BSCoopM2}. We then present results for some simple variations of the BS cooperation model, and we conclude the section with a discussion of the existing literature.


\subsection{Mode 1} 

\begin{theorem}\label{Thm:M1}
Fix the helper BS rates~$\R$, bandwidth expansion factor $\kappa$ and quantisers $\phib$.  A source/side-information tuple $(X,Y_1,Y_2,\ldots,Y_K)$  is $(\kappa,\R,\phib)$-achievable if (and only if)\footnote{For the ``only if" direction replace the \emph{strict inequality} $(*)$ with an inequality.} there exists a pmf $P_W$ on $\set{W}$ such that for all $k$ 
\begin{equation}\label{Eqn:Thm:M1}
\rt{H(X|Y_k) \step{$*$}{<} \kappa I(W;U_k) + \min\big\{ R_k,\kappa I(W;V_k|U_k) \big\},}
\end{equation}
where $(W,U_1,U_2,\ldots,U_K) \sim P_W(\cdot) T(\cdot|\cdot)$ and $V_k=\phi_k(W)$. 
\end{theorem}

Theorem~\ref{Thm:M1} is proved in Sections~\ref{Sec:Proof:Thm:M1:Con} and~\ref{Sec:Proof:Thm:M1:Ach}.


\subsection{Mode 2} 

\begin{theorem}\label{Thm:M2}
Fix the helper BS rates $\R$ and bandwidth expansion factor $\kappa$. A source/side-information tuple $(X,Y_1,Y_2,$ $\ldots,Y_K,V_1,V_2,\ldots,V_K)$ is $(\kappa,\R)$-achievable if (and only if)$^4$ there exists a pmf $P_W$ on $\set{W}$ and $K$ auxiliary random variables $(A_1,A_2,\ldots,A_K)$ such that for all $k$ we have the Markov chain $(X,Y_k) \leftrightarrow V_k \leftrightarrow A_k$,
\begin{subequations}\label{Eqn:Thm:M2}
\begin{equation}\label{Eqn:Thm:M2a}
R_k \step{$*$}{>} I(\rt{V_k};A_k|Y_k)
\end{equation}
and
\begin{equation}\label{Eqn:Thm:M2b}
H(X|A_k,Y_k) \step{$*$}{<} \rt{\kappa} I(W;U_k),
\end{equation}
\end{subequations}
where $(W,U_1,U_2,\ldots,U_K) \sim P_W(\cdot) T(\cdot|\cdot)$.
\end{theorem}

Theorem~\ref{Thm:M2} is proved in Sections~\ref{Sec:Proof:Thm:M2:Con} and~\ref{Sec:Proof:Thm:M2:Ach}.

\begin{remark}
When computing Theorem~\ref{Thm:M2}, we can assume that the alphabet of $A_k$ has a cardinality of at most $|\set{V}_k|$.
\end{remark}


\subsection{\rt{Example for Theorem~\ref{Thm:M2}}}

Consider Theorem~\ref{Thm:M2}, and choose 
$
(\rho_1,\ldots,\rho_K) \in [0,1/2]^K.
$
Suppose that the source is uniform and binary, $X \sim \text{Bern(1/2)}$; there is no receiver side information, $Y_k = \text{constant}$; and define helper BS($k$)'s side information to be
\begin{equation}\label{Eqn:Thm:M2:BinaryExampleVk}
V_k := X \oplus Z_k, \quad \quad \text{(modulo 2)},
\end{equation}
where $Z_k :=  \text{Bern}(\rho_k)$ is independent additive binary noise. The source / side-information tuple $(X,V_1,V_2,\ldots,V_K)$ is achievable if (and only if)$^4$ there exists a pmf $P_W$ on $\set{W}$ and 
$
(\alpha_1,\alpha_2,\ldots,\alpha_K) \in [0,1/2]^K
$
such that for all $k$ we have
\begin{equation}\label{Eqn:Thm:M2:BinaryExample1}
R_k \step{$*$}{>} 1 - h(\alpha_k)
\quad
\text{and}
\quad
h(\alpha_k \star \rho_k) \step{$*$}{<} \kappa I(W;U_k),
\end{equation}
where 
\begin{equation*}
h(a) := 
\left\{
\begin{array}{ll}
- a \log_2 a - (1-a)\log_2(1-a),& a \in (0,1/2],\\
0, & a = 0.
\end{array}
\right.
\end{equation*}
is the binary entropy function and 
\begin{equation*}
a \star b := a (1-b) + (1-a) b,\quad 0 \leq a,b\leq 1.
\end{equation*}

The above example is an application of Wyner's \emph{binary helper source coding problem}~\cite{Wyner-May-1975-A} (see also~\cite{Wyner-Nov-1975-A-I,Wyner-Nov-1975-A-II,Ahlswede-Nov-1975-A} and~\cite[Thm.~10.2]{El-Gamal-2011-B}). To see why~\eqref{Eqn:Thm:M2:BinaryExample1} holds, consider the following: Let $(A_1,A_2,\ldots,A_K)$ be any tuple of auxiliary random variables satisfying the conditions of Theorem~\ref{Thm:M2}. We first notice that 
\begin{multline}\label{Eqn:Thm:M2:BinaryExample2}
H(X|V_k) 
\step{a}{=} 
h(\rho_k) \\
\step{b}{\leq} 
H(X|A_k)
\step{c}{=}
H(X|A_k,Y_k) 
\step{d}{\leq}
1,\quad \forall\ k,
\end{multline}
where step (a) follows from~\eqref{Eqn:Thm:M2:BinaryExampleVk}; (b) notes that $X \leftrightarrow V_k \leftrightarrow A_k$ forms a Markov chain and applies the data processing lemma; (c) follows because $Y_k$ is a constant; and (d) follows because $X$ is binary. From~\eqref{Eqn:Thm:M2:BinaryExample2}, it follows that we can find $\alpha_k \in [0,1/2]$, for all $k$, such that 
\begin{subequations}\label{Eqn:Thm:M2:BinaryExample3}
\begin{equation}\label{Eqn:Thm:M2:BinaryExample3a}
H(X|A_k,Y_k) = h(\alpha_k \star \rho_k).
\end{equation}
In addition, we have
\begin{multline}\label{Eqn:Thm:M2:BinaryExample3b}
I(V_k;A_k|Y_k) 
\step{a}{=}
I(V_k;A_k) \\
\step{b}{=} 1 - H(V_k|A_k)
\step{c}{\geq} 
1 - h(\alpha_k).
\end{multline}
\end{subequations}
Here (a) follows because $Y_k$ is a constant, and (b) follows because $V_k \sim \text{Bern}(1/2)$ and thus $H(V_k) = 1$. Step (c) invokes Mrs Gerber's Lemma~\cite{Wyner-Nov-1975-A-I,Wyner-Nov-1975-A-II} (see also~\cite[p.~19]{El-Gamal-2011-B}) to upper bound $H(V_k|A_k)$ by $h(\alpha_k)$. More specifically, we have $X = V_k \oplus Z_k$ and $Z_k \sim \text{Bern}(\rho_k)$. Since $X \leftrightarrow V_k \leftrightarrow A_k$, it follows that
\begin{align}
\notag
& 0=I(A_k; V_k \oplus Z_k  |V_k) = I(A_k; Z_k  |V_k)\\
\label{Eqn:Thm:M2:BinaryExample4}
&\Rightarrow A_k \leftrightarrow V_k \leftrightarrow Z_k.
\end{align}
Combining~\eqref{Eqn:Thm:M2:BinaryExample4} with $I(V_k;Z_k) = 0$ shows that $Z_k$ is independent of $(V_k,A_k)$ and hence Mrs Gerber's Lemma applies.

The above discussion shows that~\eqref{Eqn:Thm:M2:BinaryExample3} holds for any choice of auxiliary random variables satisfying the conditions of Theorem~\ref{Thm:M2}. To complete the example, we need only find auxiliary random variables for which~\eqref{Eqn:Thm:M2:BinaryExample3} holds with equality. To this end, simply let $A_k$ be the output of a binary symmetric channel with input $V_k$ and crossover probability $\alpha_k$. 


\subsection{Mixed modes}

Suppose that some helper BSs have information about the codeword $\b{W}$, while others have information about the \rt{source} $\b{X}$ --- a mix of Modes 1 and 2. Let $\set{K}_1$ and $\set{K}_2$ denote the index sets of Mode 1 and 2 helper BSs respectively. It can be argued from Theorems~\ref{Thm:M1} and~\ref{Thm:M2} that a source/side-information tuple is achievable if (and only if)$^4$ there exists a pmf $P_W$ on $\set{W}$ and $|\set{K}_2|$ auxiliary random variables $\{A_k;\ k \in \set{K}_2\}$ such that $A_k \leftrightarrow V_k \leftrightarrow (X,Y_k)$ forms a Markov chain,
\begin{equation*}
H(X|Y_k) \step{$*$}{<} \rt{\kappa} I(W;U_k) + \min \{R_k, \rt{\kappa}I(W;V_k|U_k)\},\ \forall k \in \set{K}_1,
\end{equation*}
and
\begin{multline*}
R_k \step{$*$}{>} I(\rt{V}_k;A_k|Y_k)\quad \text{and}\\
H(X|A_k,Y_k) \step{$*$}{<}  \rt{\kappa}I(W;U_k),\ \forall k \in \set{K}_2.
\end{multline*}

\subsection{Broadcast capacity with helpers}

Consider Mode~1 for the bandwidth-matched case $\ns=\nc=n$, and fix a positive rate $R^*$. Suppose that there is no side information and the main BS is required to broadcast a discrete rate $R^*$ message \rt{$M$} to the receivers, \rt{where $M$} is uniformly distributed on $\{1,2,\ldots, \lfloor 2^{nR^*} \rfloor \}$. \rt{For example, in Theorem~\ref{Thm:M1} suppose that $2^{R^*}$ is an integer, $\kappa = 1$, $Y_k = \text{constant}$ and $M = \b{X}$, where $\b{X}$ is iid with a uniform distribution on $\{1,2,$ $\ldots,2^{R^*}\}$. Then $H(X|Y_k) = H(X) = R^*$ for all $k$.}

Given helper rates $\R$, we can define the \emph{helper capacity} $C(\R)$ to be the supremum of all achievable message rates~$R^*$; that is, those rates $R^*$ for which there exists a sequence of codes with vanishing probability of decoding error. It can be argued from Theorem~\ref{Thm:M1} that
\begin{multline}\label{Eqn:CompoundCapacity}
C(\R) = \max_{P_W} \min_k \big[ I(W;U_k) \\+ \min \big\{R_k, I(W;V_k|U_k)\big\} \big],
\end{multline}
where the maximisation is taken over all pmfs $P_W$ on $\set{W}$. 

If the channel outputs are defined over a common alphabet, say $\set{U}_k = \set{U}$ for all $k$, then~\eqref{Eqn:CompoundCapacity} is a type of compound channel capacity with relays. Indeed, one recovers the compound channel capacity theorem~\cite{Blackwell-Dec-1959-A,Lapidoth-May-1998-A} upon setting $R_k = 0$ in~\eqref{Eqn:CompoundCapacity}.

\subsection{Bidirectional broadcast channel with helpers} 

Consider Mode 1 with two receivers for the bandwidth matched case $\ns=\nc=n$, and fix positive rates $R_1^*$ and $R_2^*$. Recall the bidirectional setup of~\cite{Oechtering-Jan-2008-A}: The main BS has two independent uniformly distributed messages $M_1$ and $M_2$ on $\{1,2,\ldots,$ $\lfloor 2^{nR^*_1}\rfloor\}$ and $\{1,2,\ldots,\lfloor 2^{nR^*_2}\rfloor\}$ respectively; receiver 1 has $M_1$ as side information and requires $M_2$; and receiver 2 has $M_2$ as side information and requires~$M_1$. \rt{For example, in Theorem~\ref{Thm:M1} suppose that $\kappa = 1$, $2^{R^*_1}$ and $2^{R^*_2}$ are integers, $M_1 = \b{X}_1 = \b{Y}_1$ and $M_2 = \b{X}_2 = \b{Y}_2$, where $\b{X}_1$ and $\b{X}_2$ are independent with iid uniform distributions on $\{1,2,\ldots,2^{R^*_1}\}$ and $\{1,2,\ldots,2^{R^*_2}\}$ respectively. Then, setting $\b{X} = (\b{X}_1,\b{X}_2)$ gives}
\begin{align*}
\rt{H(X|Y_1)} &= \rt{H(X_1,X_2|X_1) = H(X_2) = R^*_2\quad \text{and}}\\
\rt{H(X|Y_2)} &= \rt{H(X_1,X_2|X_2) = H(X_1) = R^*_1.} 
\end{align*}
For fixed helper rates $(R_1,R_2)$, we can define the \emph{helper capacity region} $\set{C}(R_1,R_2)$ to be closure of the set of all $(R_1,R_2)$-achievable rate pairs $(R^*_1,R^*_2)$. It can be argued from Theorem~\ref{Thm:M1} that $\set{C}(R_1,R_2)$ is equal to the set of all $(R^*_1,R^*_2)$ for which there exists a pmf $P_W$ on $\set{W}$ such that 
\begin{align*}
R^*_1 &\leq I(W;U_2) + \min \big\{R_2, I(W;V_2|U_2) \big\}\\
R^*_2 &\leq I(W;U_1) + \min \big\{R_1,  I(W;V_1|U_1) \big\}.
\end{align*}  


\subsection{Other work \& Operational source-channel separation} 

Consider Mode 1 and Theorem~\ref{Thm:M1}. If the helper rates are all set to zero, then~\eqref{Eqn:Thm:M1} becomes
\begin{equation}\label{Eqn:Compare1}
H(X|Y_k) \step{$*$}{<}  \rt{\kappa} I(W;U_k), \quad \forall\ k,
\end{equation}
and we recover the setup of~\eqref{Eqn:Tuncel}. Now suppose that for a given pmf $P_W$ and scalar quantisers $\phib$ we have $R_k >  \rt{\kappa}H(V_k|U_k)$ for all $k$. If $(\b{V}_k,\b{U}_k)$ behaves like a discrete memoryless source, then BS($k$) can reliably send $\b{V}_k$ to receiver $k$ using a Slepian-Wolf code of rate $R_k$~\cite{Slepian-Jul-1973-A}. The receiver effectively has the combined channel output $(\b{U}_k,\b{V}_k)$. Since \eqref{Eqn:Thm:M1} simplifies to 
\begin{equation*}
H(X|Y_k) \step{$*$}{<} \rt{\kappa} I(W;U_k,V_k), \quad \forall\ k,
\end{equation*}
we again return to the result in~\eqref{Eqn:Tuncel}, where the $k$-th channel output $U_k$ is replaced by $(U_k,V_k)$. 

For other helper rates, we note the similarity of~\eqref{Eqn:Thm:M1} to Kim's capacity theorem~\cite[Thm.~1]{Kim-Mar-2008-A} for the primitive relay channel. Intuitively, the right hand side of~\eqref{Eqn:Thm:M1} is the maximum rate at which information can be sent to receiver~$k$.  \rt{This intuition, however, should \rt{treated with care} because, for example, the classical Shannon approach of \emph{strictly} separating source and channel coding is suboptimal\footnote{To see why \emph{strict} source-channel separation fails, set $R_k = 0$ for all $k$ and consider the examples in~\cite{Tuncel-Apr-2006-A}.}. Nonetheless, it is natural to wonder whether Kim's simple timesharing proof of~\cite[Thm.~1]{Kim-Mar-2008-A} can be modified to prove Theorem~\ref{Thm:M1}. While we do not take the timesharing approach in this paper, D.~G\"{u}nd\"{u}z has noticed that it may indeed be possible to give such a proof of Theorem~\ref{Thm:M1} using the \emph{semiregular encoding} and \emph{backward decoding} techniques developed in~\cite[App.~B]{Gunduz-Apr-2013-A} (these techniques, for example, give an alternative proof of the no-cooperation case shown in~\eqref{Eqn:Tuncel}).}  

\rt{
The single-letter characterisations in Theorems~\ref{Thm:M1} and~\ref{Thm:M2} depend only on the marginal source and channel distributions, instead of the complete joint source-channel distribution\footnote{All of the entropy and mutual information functions in Theorems~\ref{Thm:M1} and~\ref{Thm:M2} depend on either the source variables or the channel variables, but not both.} --- the latter being more typical in the joint source-channel coding literature, e.g., see~\cite{Cover-Nov-1980-A}. The separation of source and channel variables in Theorems~\ref{Thm:M1} and~\ref{Thm:M2}  is reminiscent of \emph{operational separation} described in~\cite{Tuncel-Apr-2006-A} and can be similarly understood\footnote{More detailed discussions on the various types of source-channel separation can be found in~\cite{Tuncel-Apr-2006-A,Gunduz-Sep-2009-A,Gunduz-Apr-2013-A}.}. Indeed, in both modes we will see that it is optimal to separate the source, channel and helper codebooks as well as the encoders, but joint decoding across all three codebooks is required. In particular, the approach taken in this paper is to first require that receiver $k$ decodes a list of likely source sequences using a joint source-channel decoder on its channel output $\b{U}_k$ and side information $\b{Y}_k$. The receiver then determines the correct source sequence, within this list, using the helper BS's message and codebook. The list decoding approach is particularly useful because it highlights an operation duality between Modes 1 and 2: The helper BS's task in both modes is to help the receiver resolve the correct source sequence from the receiver's list.
}

\rt{The problem setup of Mode $2$ is a special case of the more general problem considered by G\"{u}nd\"{u}z, Erkip, Goldsmith and Poor in~\cite[Sec.~V]{Gunduz-Apr-2013-A}. G\"{u}nd\"{u}z \emph{et al.}~presented an achievability result for the general problem in~\cite[Thm.~3]{Gunduz-Apr-2013-A}, however, this result is not optimal in the case of Theorem~\ref{Thm:M2}.} In this paper, helper BS($k$) provides information directly about the \rt{source} $\b{X}$ via a `quantised' version of $\b{V}_k$. The quantisation is specified by the auxiliary random variable $A_k$ in a similar way to the quantisation in Wyner's helper side-information problem~\cite{Wyner-May-1975-A}, \cite[p.~575]{Cover-2006-B} or the Wyner-Ziv rate-distortion problem~\cite{Wyner-Jan-1976-A}.

Comparing Theorems~\ref{Thm:M1} and~\ref{Thm:M2}: Increasing the helper rates in Theorem~\ref{Thm:M2} allows larger `quantisation rates' and reductions in the left hand side of~\eqref{Eqn:Thm:M2b}. In contrast, increasing the helper rates in Theorem~\ref{Thm:M1} improves the `relay capacity' and increases the right hand side of~\eqref{Eqn:Thm:M1}. We will see that these properties are dual consequences of the same random-coding idea.

Finally, we note that Theorems~\ref{Thm:M1} and~\ref{Thm:M2} are existential statements that do not give constructive arguments for low-complexity codes. That being said, however, the single-letter expressions and (as we will see) the structure of the random-coding achievability proofs give insight into the architecture of good low-complexity codes. For example, the hash-and-forward  technique used in Mode~1 is similar to distributed source coding using LDPC codes~\cite{Chen-Sep-2009-A}. Similarly, in Mode~2, preliminary work suggests that (nonlinear) trellis codes and rate-distortion codes perform well for quantising $\b{V}_k$~\cite{Morshed-Feb-2014-C}. Finally, recent work~\cite{Iscan-Feb-2015-C} suggests that repeat-accumulate codes can be useful for Slepian-Wolf coding over broadcast channels.


\section{Slepian-Wolf Coding over Broadcast Channels with List decoding}\label{Sec:LD}

It is useful to consider a list-decoding extension to~\eqref{Eqn:Tuncel} before proving Theorems~\ref{Thm:M1} and~\ref{Thm:M2}. In this section, suppose that there is \emph{no} BS cooperation and the receivers employ list decoding. 


\subsection{Setup and Main Result}

Let $\Omega(L) := \{ \set{L} \subseteq \set{X}^{\rt{\ns}} : |\set{L}| = L \}$ denote the collection of all subsets of $\set{X}^{\rt{\ns}}$ with cardinality $L$. An $(\rt{\ns,\nc},L_1,L_2,\ldots,$ $L_K)$ \emph{list code} is a collection of $(K+1)$ maps $(f,g_1,g_2,$ $\ldots,g_K)$, where 
\begin{equation*}
f : \rt{\set{X}^\ns} \longrightarrow \rt{\set{W}^{\nc}}
\end{equation*}
is the encoder at the transmitter and 
\begin{equation*}
g_k : \rt{\set{U}_k^{\nc}} \times \rt{\set{Y}_k^{\ns}} \longrightarrow \Omega(L_k)
\end{equation*}
is the list decoder at receiver $k$. Upon observing the channel output $\b{U}_k$ and side information $\b{Y}_k$,  receiver $k$ computes the list
\begin{equation*}
\set{L}_k := g_k(\b{U}_k,\b{Y}_k).
\end{equation*}
An error is declared at receiver $k$ if $\b{X} \notin \set{L}_k$. 

If~\eqref{Eqn:Tuncel} holds, then~\cite[Thm.~6]{Tuncel-Apr-2006-A} guarantees the existence of a sequence of list codes with $|\set{L}_k| = 1$ and $\Pr[\b{X} \notin \set{L}_k] \rightarrow 0$ for all $k$. On the other hand: If \eqref{Eqn:Tuncel} does not hold, then $|\set{L}_k|$ must grow exponentially in \rt{$\ns$} to ensure $\Pr[\b{X} \notin \set{L}_k] \rightarrow 0$. We are concerned with the smallest such exponent.
 
\begin{definition}\label{Def:List}
\rt{Fix the bandwidth expansion factor $\kappa$ and} list exponents $\D = (D_1,D_2,\ldots,D_K)$, with $D_k \geq 0$, $\forall k$. We say that the pair \rt{$(\kappa,\D)$} is \emph{achievable} if for any $\epsilon > 0$ there exists a $(\rt{\ns},$ $\rt{\nc},L_1,L_2,\ldots,$ $L_K)$ list code such that 
\begin{subequations}\label{Eqn:DefList}
\begin{equation}\label{Eqn:DefLista}
\rt{\frac{\nc}{\ns} = \kappa,}
\end{equation}
\begin{equation}\label{Eqn:DefListb}
L_k \leq 2^{\rt{\ns} D_k}\quad \text{and}\quad \Pr\left[\b{X} \notin \set{L}_k \right] \leq \epsilon,\ \forall\ k,
\end{equation}
\end{subequations}
\rt{where $\ns$ and $\nc$ are sufficiently large.}
\end{definition}

\rt{The next lemma is proved in Sections~\ref{Sec:Proof:Lem:SW-List:Con} and~\ref{Sec:Proof:Lem:SW-List:Ach}.}

\begin{lemma}\label{Lem:SW-List}
\rt{$(\kappa,\D)$} is achievable if (and only if)\footnote{Replace the \emph{strict inequality} ($*$) with an inequality.} there exists a pmf $P_W$ on $\set{W}$ such that 
\begin{equation*}
D_k \step{$*$}{>} \max\big\{ H(X|Y_k)  - \rt{\kappa} I(W;U_k),\ 0 \big\}, \quad \forall\ k,
\end{equation*}
where $(W,U_1,U_2,\ldots,U_K) \sim P_W(\cdot) T(\cdot|\cdot)$.
\end{lemma}

Lemma~\ref{Lem:SW-List} is quite intuitive: The best exponent of receiver~$k$'s list size can be larger, but not smaller, than the equivocation in $X$ given $Y_k$ minus the information conveyed over the channel. 

\begin{remark}
Definition~\ref{Def:List} is a lossy generalisation of the setup for~\eqref{Eqn:Tuncel}. The standard (per-letter / average distortion) generalisation of~\eqref{Eqn:Tuncel} is called \emph{``Wyner-Ziv Coding over broadcast channels''}~\cite{Nayak-Apr-2010-A1}, and it is a formidable open problem that includes Heegard and Berger's rate-distortion function~\cite{Heegard-Nov-1985-A,Timo-Aug-2011-A,Timo-Nov-2010-A} as well as the broadcast capacity region~\cite{El-Gamal-2011-B}.
\end{remark}

\begin{remark}
Definition~\ref{Def:List} and Lemma~\ref{Lem:SW-List} are related to Chia's recent list-decoding result~\cite[Prop.~1]{Chia-Jul-2014-C} for Heegard and Berger's rate-distortion problem~\cite{Heegard-Nov-1985-A}. For example, suppose that $\kappa = 1$ and we replace the memoryless BC $T(\cdot|\cdot)$ in our model with a noiseless source-coding `index' channel, with alphabet $\{1,2,\ldots,$ $\lfloor \rt{2^{\ns R_\text{s}}} \rfloor\}$. In this case, the mutual information $I(W;U_k)$ transforms to the source-coding rate $R_\text{s}$ and Lemma~\ref{Lem:SW-List} reduces to~\cite[Prop.~1]{Chia-Jul-2014-C} 
\begin{equation*}
R_\text{s} \step{$*$}{>} \max_k \big\{ H(X|Y_k) - D_k \big\}.
\end{equation*}    
\end{remark}

\begin{remark}\label{Rem:ListVTuncel}
Lemma~\ref{Lem:SW-List} is consistent with Tuncel's result for unique decoding~\eqref{Eqn:Tuncel} in the following sense. Suppose that we are interested in unique decoding and hence the all-zero list exponent vector $\D = (0,0,\ldots,0)$. The reverse (converse) assertion of Lemma~\ref{Lem:SW-List} shows that $(\kappa,\D)$ is achievable \emph{only if}
\begin{equation}\label{Eqn:Rem:ListVTuncel}
H(X|Y_k) \leq \kappa I(W;U_k),\quad \forall k. 
\end{equation}
The forward (achievability) assertion of Lemma~\ref{Lem:SW-List}, unfortunately, does not include the all-zero list exponent. It does, however, say the following: Any arbitrarily small positive list exponent $\D$ is achievable if~\eqref{Eqn:Rem:ListVTuncel} holds.  
\end{remark}

\begin{remark}\label{Rem:ListANDTuncel}
It is natural to combine and extend~\eqref{Eqn:Tuncel} and Lemma~\ref{Lem:SW-List} as follows: Suppose that a subset 
\begin{equation*}
\set{K}_\text{List} \subseteq \{1,2,\ldots,K\}
\end{equation*}
of receivers employ list decoding as in~\eqref{Eqn:DefListb}, and the remaining receivers 
\begin{equation*}
\set{K}_\text{Unique} := \{1,2,\ldots,K\} \backslash \set{K}_\text{List}
\end{equation*}
employ unique decoding as in $\Pr[\hat{\b{X}}_k \neq \b{X}] \leq \epsilon$ (for example, see~\cite{Tuncel-Apr-2006-A}). The bandwidth expansion factor $\kappa$ is jointly achievable with list exponents $\{D_k \geq 0;\ k \in \set{K}_\text{List}\}$ for the receivers in $\set{K}_\text{List}$ and unique decoding for receivers in $\set{K}_\text{Unique}$ if (and only if)$^4$ there exists a pmf $P_W$ on $\set{W}$ such that 
\begin{subequations}\label{Eqn:Rem:ListANDTuncel}
\begin{equation}
D_k \step{$*$}{>} \max\big\{H(X|Y_k) - \kappa I(W;U_k),\ 0\big\},\ \forall\ k \in \set{K}_\text{List}
\end{equation}
and
\begin{equation}
H(X|Y_k) \step{$*$}{<} \kappa I(W;U_k),\quad \forall\ k \in \set{K}_\text{Unique},
\end{equation}
\end{subequations}
where $(W,U_1,U_2,\ldots,U_K) \sim P_W(\cdot) T(\cdot|\cdot)$. The reverse (converse) assertion of~\eqref{Eqn:Rem:ListANDTuncel} automatically follows from Lemma~\ref{Lem:SW-List} upon setting $D_k = 0$ for all $k \in \set{K}_\text{Unique}$. As to the forward (achievability) assertion: The random codebook used in~\cite[Thm.~6]{Tuncel-Apr-2006-A} has the same structure as that used to prove Lemma~\ref{Lem:SW-List}, so we need only combine the error analysis in Section~\ref{Sec:Proof:Lem:SW-List:Ach} with the analysis in~\cite[Sec.~IV]{Tuncel-Apr-2006-A} using, say, the union bound. We omit the details. 
\end{remark}


\subsection{Discussion: List Decoding and the Operational Duality of Theorems~\ref{Thm:M1} and~\ref{Thm:M2}}

\rt{It turns out that the following approach to BS cooperation is optimal in both modes: Use a good list code on the broadcast channel, and task BS($k$) with helping receiver~$k$ determine which element of its decoded list $\set{L}_k$ is equal to the source~$\b{X}$. This list will, with high probability, include $\b{X}$ and have $|\set{L}_k| \approx 2^{\ns D_k}$ elements. To resolve receiver~$k$'s uncertainty, BS$(k$) needs to encode its side information $\b{V}_k$ at a rate $R_k$ that is proportional to the list exponent $D_k$. In both modes, the smallest achievable rate $R_k$ is fundamentally determined by Lemma~\ref{Lem:SW-List}. Theorems~\ref{Thm:M1} and~\ref{Thm:M2} are duals in the operational sense that changing from Mode $1$ to Mode $2$ (or, vice versa) does not change the underlying coding problem --- it only changes BS($k$)'s approach to the problem. The side information $\b{V}_k$ in Mode $2$ is directly correlated with the source $\b{X}$, and, in this setting, it is optimal for BS($k$) to use a good source code from Wyner's `helper' source coding problem~\cite{Wyner-May-1975-A}. In Mode $1$, on the other hand, the side information $\b{V}_k$ is a scalar quantised version of the channel codeword, and it is optimal for BS($k$) to use a version of Kim's `random-hashing' for the relay channel~\cite{Kim-Mar-2008-A}.} The remainder of the paper is devoted to proving Lemma~\ref{Lem:SW-List} and Theorems~\ref{Thm:M1} and~\ref{Thm:M2}. 


\section{Proof of Lemma~\ref{Lem:SW-List} --- Converse}\label{Sec:Proof:Lem:SW-List:Con}

Fix $\epsilon > 0$, and suppose that we have a $(\rt{\ns},\rt{\nc},L_1,L_2,$ $\ldots,L_K)$ list code such that~\eqref{Eqn:DefList} holds.  As before, let $\b{W} = f(\b{X})$ and $\b{U}_k = (U_{k,1},$ $U_{k,2},\ldots,\rt{U_{k, \nc}})$ denote the transmitted codeword and the channel output at receiver~$k$. 

The first step mirrors that of~\cite[Thm.~6]{Tuncel-Apr-2006-A}. Consider the $j$-th symbol $W_j$ of $\b{W} = (W_1,W_2,\ldots,$ $\rt{W_\nc})$, and let $P_{W_j}$ denote its pmf. Construct a \emph{timeshared} pmf $P_{\tilde{W}}$ on $\set{W}$ by setting
\begin{equation}\label{Eqn:TimeshareCon}
P_{\tilde{W}}(w)
:= 
\frac{1}{\rt{\nc}} \sum_{j = 1}^{\rt{\nc}} P_{W_j}(w),\quad w \in \set{W}.
\end{equation}

Let $(\tilde{W},\tilde{U}_1,\tilde{U}_2,\ldots,\tilde{U}_K) \sim P_{\tilde{W}}(\cdot) T(\cdot |\cdot)$. We have
\begin{align}
\rt{\nc} I(\tilde{W};\tilde{U}_k) 
\notag
&\step{a}{\geq} 
\sum_{i = 1}^{\rt{\nc}} I(W_i;U_{k,i}) \\
\notag
&\step{b}{\geq} 
I(\b{W};\b{U}_k) \\
\notag
&\step{c}{\geq}
I(\b{X};\b{U}_k|\b{Y}_k) \\
\label{Eqn:Converse1}
&\step{d}{=}
 \rt{\ns}H(X | Y_k) - H(\b{X} | \b{Y}_k,\b{U}_k).
\end{align}
Notes: (a) use Jensen's inequality~\cite[Thm.~2.7.4]{Cover-2006-B}; (b) \rt{
\begin{align*}
&\sum_{i = 1}^\nc I(W_i;U_{k,i})\\
&\quad= 
\sum_{i = 1}^\nc 
\Big(H(U_{k,i}) - H(U_{k,i}|W_i)\Big)\\
&\quad\geq 
H(\b{U}_k) - \sum_{i = 1}^\nc  H(U_{k,i}|W_i)\\
&\quad\step{*}{=} 
H(\b{U}_k) - \sum_{i = 1}^\nc  H(U_{k,i}|\b{W},U_{k,1},U_{k,2},\ldots,U_{k,i-1})\\
&\quad= 
H(\b{U}_k) - H(\b{U}_k|\b{W}),
\end{align*}
where ($*$) follows because the broadcast channel is memoryless and therefore $U_{k,i} \leftrightarrow W_i \leftrightarrow (\b{W},U_{ k,1},U_{ k,2},\ldots,U_{ k,i-1})$
forms a Markov chain;} (c) $(\b{X},\b{Y}_k) \leftrightarrow \b{W} \leftrightarrow \b{U}_k$ forms a Markov chain; and (d) the source is iid.

We now use a list-decoding version of Fano's inequality, e.g., see~\cite[Lem.~1]{Chia-Jul-2014-C} or~\cite[Lem.~1]{Kim-May-2008-A}:
\begin{multline*}
H(\b{X}|\b{Y}_k,\b{U}_k) \leq \log|\set{L}_k| + 1 \\
+(\rt{\ns} \log|\set{X}| - \log|\set{L}_k|)
\Pr\left[\b{X} \notin \bigcap_{k = 1}^K \set{L}_k\right].
\end{multline*}
By~\eqref{Eqn:DefListb}, and since for any $k$, $\Pr[\b{X} \notin \set{L}_k] \geq \Pr[\b{X} \notin \cap_{k'} \set{L}_{k'}]$, this inequality implies 
\begin{align}
H(\b{X}|\b{Y}_k,\b{U}_k) 
\label{Eqn:Fano}
&\leq \rt{\ns}\big(D_k + \varepsilon(\rt{\ns},\epsilon)\big),
\end{align} 
where 
\begin{equation*}
\varepsilon(\rt{\ns},\epsilon) := \frac{1}{\rt{\ns}} + \epsilon(\rt{1} + \log|\set{X}| - D_k - \rt{\epsilon}).
\end{equation*}
Combining~\eqref{Eqn:DefLista}, \eqref{Eqn:Converse1} and~\eqref{Eqn:Fano}, we have 
\begin{equation*}
\rt{\kappa} I(\tilde{W};U_k) \geq H(X | Y_k) - D_k - \varepsilon(\rt{\ns},\epsilon).
\end{equation*}

To complete the converse: Take any positive and vanishing sequence $\{\epsilon\} \rightarrow 0$. Consider the corresponding sequence of list codes (with increasing blocklengths \rt{$\ns$ and $\nc$}) and time-shared pmfs $\{P_{\tilde{W}}\}$. Since $\set{W}$ is a finite alphabet, by the Bolzano-Weierstrass theorem, $\{P_{\tilde{W}}\}$ will contain a convergent subsequence with respect to the variational distance. Let $P^*_{\tilde{W}}$ denote the limit of the convergent subsequence and $\tilde{W}^* \sim P^*_{\tilde{W}}$. We then have $\rt{\kappa} I(\tilde{W}^*;U_k) \geq H(X | Y_k) - D_k$ by the continuity of mutual information~\cite[Sec.~2.3]{Yeung-2008-B}. \hfill $\blacksquare$


\section{Proof of Lemma~\ref{Lem:SW-List} --- Achievability}\label{Sec:Proof:Lem:SW-List:Ach}


\rt{We restrict attention to the bandwidth matched case ($\kappa = 1$ and $\ns = \nc = n$), to help simplify notation and elucidate the main ideas of the achievability proof. Extending this proof to the bandwidth mismatched case is relatively easy, because we will use separate source and channel codebooks and the error probability bounds depend only on the marginal source and channel distributions.}

\subsection{Notation and Letter-Typical Sets}

\rt{
For any given random variable $\omega$ and set $\Omega$, let us denote the indicator function for the event that $\omega$ falls in $\Omega$ by 
\begin{equation*}
\indicator{\omega \in \Omega } := \left\{ 
\begin{array}{ll}
1 &\text{ if } \omega \in \Omega\\
0 &\text{ otherwise.} 
\end{array}
\right.
\end{equation*}
} 

The proof will use \emph{letter typical} sets~\cite{Kramer-2008-A}. Consider a pair of random variables $(A,B) \sim P_{A,B}$ on $\set{A} \times \set{B}$, where $\set{A}$ and $\set{B}$ are finite alphabets. Let $P_A$ denote the marginal pmf of $A$. For $\es > 0$ and a positive integer $n$, the \emph{typical set} of $P_A$ is 
\begin{multline*}
\set{T}_\epsilon^n(P_A) := \Big\{ \b{a} \in \set{A}^n : \\ \Big| \frac{1}{n} N(a'|\b{a}) - P_A(a') \Big| \leq \epsilon P_A(a'),\ \forall a' \in \set{A} \Big\},
\end{multline*}
where $N(a'|\b{a})$ represents the number of occurrences of $a'$ in the sequence $\b{a}$. The \emph{jointly typical set} of $P_{A,B}$ is 
\begin{multline*}
\set{T}^n_\epsilon(P_{A,B}) := \Big\{(\b{a},\b{b}) \in \set{A}^n \times \set{B}^n :
\Big| \frac{1}{n} N(a',b'|\b{a},\b{b})\\ 
- P_{A,B}(a',b') \Big| 
\leq \epsilon P_{A,B}(a',b'),\ \forall (a',b') \Big\}.
\end{multline*}
The \emph{conditionally typical set} of $P_{A,B}$ given $\b{b} \in \set{B}^n$ is 
\begin{equation*}
\set{T}^n_\es(P_{A,B}|\b{b}) := 
\big\{ \b{a} \in \set{A}^n : 
(\b{a},\b{b}) \in \set{T}^n_\es(P_{A,B}) \big\}.
\end{equation*}
The proof will frequently use the property that joint typicality implies marginal typicality,
\begin{equation*}
(\b{a},\b{b}) \in \set{T}^n_\epsilon(P_{A,B}) \Rightarrow \b{a} \in \set{T}^n_\epsilon(P_{A})
\text{ and } 
\b{b} \in \set{T}^n_\epsilon(P_{B}),
\end{equation*}
and the following lemmas. Let
\begin{equation*}
\mu_{A}  := \min_{a \in \text{supp}(P_{A})} P_{A}(a),
\end{equation*}
and
\begin{equation*}
\mu_{A,B} := \min_{(a,b) \in \text{supp}(P_{A,B})} P_{A,B}(a,b),
\end{equation*} 
where $\text{supp}(\cdot)$ denotes the support set of the indicated distribution.

\begin{lemma}\label{Lem:Kramer-1}
If $\b{A} := (A_1,A_2,\ldots,A_n)$ is generated iid with $P_A$, $0 < \es \leq \mu_A$ and $\b{a} \in \set{T}^n_\es(P_A)$, then~\cite[Thm.~1.1]{Kramer-2008-A}
\begin{equation*}
2^{-nH(A)(1+\es)} \leq \Pr[\b{A} = \b{a}] \leq 2^{-nH(A)(1-\es)}
\end{equation*}
and
\begin{equation*}
1 - 2|\set{A}|\exp(-n \es^2\mu_A) \leq \Pr[\b{A} \in \set{T}^n_\es(P_A)] \leq 1.
\end{equation*}
\end{lemma}

\begin{lemma}\label{Lem:Kramer-2}
If $\b{A} := (A_1,A_2,\ldots,A_n)$ is generated iid with $P_A$, $0 < \es_1 < \es \leq \mu_{AB}$ and $\b{b} \in \set{T}^n_{\es_1}(P_B)$, then~\cite[Thm.~1.3]{Kramer-2008-A}
\begin{equation*}
\Pr\big[\b{A} \in \set{T}^n_{\es}(P_{A,B}|\b{b})\big] 
\leq
2^{-n(I(A;B)-2\es H(A))}
\end{equation*}
and
\begin{equation*}
\Pr\big[\b{A} \in \set{T}^n_{\es}(P_{A,B}|\b{b})\big] 
\geq (1-\zeta_n) 2^{-n(I(A;B)+2 \es H(A))},
\end{equation*}
where\footnote{Here we use I.~Sason's correction to~\cite[Thm.~1.3]{Kramer-2008-A}, see~\cite[pp.~140--154]{Kramer-2012-L}.}
\begin{equation*}
\zeta_n := 2 |\set{A}| |\set{B}| \exp \left( -2n(1-\es_1)\ \frac{(\es - \es_1)^2}{1 + \es_1}\ \mu^2_{AB} \right).
\end{equation*}
\end{lemma}


\subsection{Distributions and Typicality Constants}

Pick any pmf $P_W$ on $\set{W}$. Let 
\begin{equation*}
X \sim P_X, \quad (X,Y_k) \sim P_{X,Y_k} \quad \text{and}\quad (W,U_k) \sim P_{W,U_k}
\end{equation*}
denote the pmfs of the indicated variables. Fix any arbitrarily small constants $\es,\es_1,\ec$ and $\ec_1$ satisfying 
\begin{subequations}\label{Eqn:epsilons}
\begin{multline}
0 < \ec_1 < \ec < \min_k \mu_{W,U_k} \\
\ \ \text{ and }
\ \ 0 < \es_1 < \es < \min_{k} \mu_{X,Y_k},
\end{multline}
with
\begin{equation}\label{Eqn:AlphaTest}
\es < \frac{ \min_k \mu_{W,U_k}}{2 H(X) \ln 2} \ec^2.
\end{equation}
\end{subequations}


\subsection{Code Construction and Encoding}\label{Sec:ListEncoder}

The encoder mirrors that of~\cite[Thm.~6]{Tuncel-Apr-2006-A}. Randomly generate a source codebook $\set{C}_X$ with 
\begin{equation}\label{Eqn:M}
M = \lfloor 2^{n H(X) (1 + \es)} \rfloor
\end{equation}
codewords, each of length $n$, by selecting symbols from $\set{X}$ in an iid fashion using $P_X$:
\begin{equation*}
\set{C}_X := \Big\{ \b{X}(m) = \big(X_1(m),X_2(m),\ldots,X_n(m)\big) \Big\}_{m = 1}^{M}.
\end{equation*}
In the same way, generate a channel codebook $\set{C}_W$ with $M$ codewords of length $ n$ using $P_W$:
\begin{equation*}
\set{C}_W := \Big\{ \b{W}(m) = \big(W_1(m),W_2(m),\ldots,W_{ n}(m)\big) \Big\}_{m = 1}^{M}.
\end{equation*}

Upon observing the \rt{source} $\b{X}$, the transmitter searches through the source codebook $\set{C}_X$ for the smallest index $m$ such that $\b{X} = \b{X}(m)$.  If successful, the transmitter sends the corresponding channel codeword $\b{W}(m)$; and, if unsuccessful, it sends $\b{W}$ generated iid $\sim P_W$.


\begin{figure*}[b!]
\hrule
\setcounter{equation}{27}
\input{./DoubleColumnEqns/Eqn-AchProof-Step3aa}
\setcounter{equation}{19}
\end{figure*}

\subsection{List Decoding at Receiver $k$}\label{Sec:ListDecoder}

The decoder (and error analysis) differ from~\cite[Thm.~6]{Tuncel-Apr-2006-A}. Upon observing the channel output $\b{U}_k$ and side information $\b{Y}_k$, receiver $k$ outputs the list
\begin{multline}\label{Eqn:List}
\set{L}_k := \Big\{ \b{X}(m) \in \set{C}_X  : 
\big(\b{X}(m),\b{Y}_k\big) \in \set{T}^n_\es(P_{X,Y_k})\\
\text{ and }\big(\b{W}(m),\b{U}_k\big) \in \set{T}_\ec^n(P_{W,U_k}) \Big\}.
\end{multline}
An error is declared at receiver $k$ if the \rt{source} is not in the list $\b{X} \notin \set{L}_k$ or the list is too large\begin{equation*}
|\set{L}_k| > 2^{nD_k}.
\end{equation*}


\subsection{Error Analysis: \rt{Decoding error event $\set{E}$}}

Denote the event of an error at any receiver by 
\begin{equation}\label{Eqn:LDError}
\set{E} :=
\bigcup_{k=1}^K \Big(
\{\b{X} \notin \set{L}_k \big\}
\cup 
\big\{ |\set{L}_k | > 2^{nD_k} \big\} \Big).
\end{equation}
By the union bound,
\begin{equation}
\label{Eqn:AchProof:UnionBnd}
\Pr[\set{E}] 
\leq \sum_{k = 1}^K 
\Big(\Pr\big[ \b{X} \notin \set{L}_k \big] 
+ \Pr\big[ |\set{L}_k | > 2^{nD_k} \big]
\Big).
\end{equation} 

In the following subsections, we show that the average error probability $\Pr[\set{E}]$ satisfies
\begin{equation}\label{Eqn:Error:AchProof}
\Pr[\set{E}] 
\leq 
b\ 2^{-a n},
\end{equation}
for some finite $a,b > 0$, whenever $\es$ and $\ec$ satisfy~\eqref{Eqn:epsilons} and 
\begin{equation*}
D_k > \max\big\{H(X|Y_k) -  I(W;U_k),0\big\}, \quad \forall\ k.
\end{equation*}
Therefore, for any $\epsilon^* > 0$ there exists an $(n,L_1,L_2,\ldots,L_K)$ list code such that $\Pr[\b{X} \notin \set{L}_k] \leq \es^*$ and $|\set{L}_k| \leq 2^{nD_k}$ for all~$k$. 

The remainder of this section is devoted to proving~\eqref{Eqn:Error:AchProof}. The derivation of the bound is a little tedious and the reader needs only~\eqref{Eqn:Error:AchProof} to proceed to the achievability proofs of Theorems~\ref{Thm:M1} and~\ref{Thm:M2} in Sections~\ref{Sec:Proof:Thm:M1:Ach} and~\ref{Sec:Proof:Thm:M2:Ach} respectively.


\subsection{Error Analysis: \rt{Probability $\b{X}$ is not in the source codebook}} 

The probability that the source is not in the source codebook $\Pr[\b{X} \notin \set{C}_X]$ is bounded by
\begin{equation}\label{Eqn:AchProof:Step1} 
\Pr[\b{X} \notin \set{C}_X] \leq b_1\ 2^{-a_1 n},
\end{equation}
where 
\begin{equation*}
a_1:=\min\big\{\es_1^2\mu_X,(H(X)\cdot(\es-\es_1))\big\}/\ln 2
\end{equation*}
and $b_1:=2|\set{X}|+1$ are both positive by~\eqref{Eqn:epsilons}.

The steps leading to~\eqref{Eqn:AchProof:Step1} are 
\begin{align}
\notag
&\Pr[\b{X} \notin \set{C}_X]\\
\notag
&\leq 
\Pr\big[ \b{X} \notin \set{T}^n_{\es_1}(P_X)\big]
+ 
\Pr\big[ \b{X} \notin \set{C}_X  \big| \b{X} \in \set{T}^n_{\es_1}(P_X)\big]\\
\notag
&\step{a}{\leq} 
2 |\set{X}| e^{-n \es_1^2\mu_X}
+ \Pr\Bigg[\bigcap_{m = 1}^M \{\b{X}(m) \neq \b{X}\} \Bigg| \b{X} \in \set{T}^n_{\es_1}(P_X)\Bigg]\\
\notag
&\step{b}{=} 
2 |\set{X}| e^{-n \es_1^2\mu_X}
+ \prod_{m = 1}^M \Big(1 - \Pr\big[ \b{X}(m) = \b{X} \big| \b{X} \in \set{T}^n_{\es_1}(P_X)\big] \Big)\\
\notag
&\step{c}{\leq} 
2 |\set{X}| e^{-n \es_1^2\mu_X} 
+ \Big(1 - 2^{-n H(X) (1 + \es_1)} \Big)^M \\
\notag 
&\step{d}{\leq} 
2 |\set{X}| e^{-n \es_1^2\mu_X}
+ \exp\big(-M 2^{- n H(X)(1 + \es_1)}\big) \\ 
\label{Eqn:AchProof:Step1a}
&\step{e}{\leq} 
2 |\set{X}| e^{-n \es_1^2\mu_X} + \exp(-2^{n H(X) (\es - \es_1)}).
\end{align}
Notes: 
\begin{itemize}
\item[a.] apply Lemma~\ref{Lem:Kramer-1};
\item[b.] the codewords in $\set{C}_X$ are independent;
\item[c.] use Lemma~\ref{Lem:Kramer-1} with $\b{X}(m)$ iid $\sim P_X$;
\item[d.] use the inequality 
\begin{equation*}
(1-c)^M \leq e^{-cM},\quad \forall\ \text{$M \geq 1$,  $c \in [0,1]$; and}
\end{equation*}
\item[e.] bound $M$ via~\eqref{Eqn:M}.
\end{itemize}
The bound in~\eqref{Eqn:AchProof:Step1} follows since $H(X)(\es-\es_1) > 0$ from~\eqref{Eqn:epsilons}.

\subsection{Error Analysis: \rt{Probability $\b{X}$ is not in receiver~$k$'s list $\set{L}_k$}}

Consider the probability that the \rt{source} $\b{X}$ is not in receiver~$k$'s list $\set{L}_k$. We have 
\begin{multline}\label{Eqn:AchProof:Step2}
\Pr\big[\b{X} \notin \set{L}_k \big]
\leq \Pr[\b{X} \notin \set{C}_X] \\
+ \Pr[(\b{X},\b{Y}_k) \notin \set{T}_{\es_1}(P_{X,Y_k})]
+\Pr\big[{\set{S}_1} \big],
\end{multline} 
where 
\begin{equation*}
{\set{S}_1} := \{\b{X} \notin \set{L}_k\} \cap \{(\b{X},\b{Y}_k) \in \set{T}^n_{\es_1}(P_{XY_k})\} \cap \{\b{X} \in \set{C}_X \}.
\end{equation*}

The probability $\Pr[{\set{S}_1}]$ is bounded from above by 
\begin{equation}\label{Eqn:AchProof:Step3a}
\Pr\big[{\set{S}_1}\big] \leq 2 |\set{W}| |\set{U}_k|\ 2^{-a_2 n}
\end{equation}
where 
\begin{equation*}
a_2 := \delta^2\mu_{W,U_k}/\ln 2 -2\es_1H(X)
\end{equation*}
is positive by~\eqref{Eqn:epsilons}. The steps leading to~\eqref{Eqn:AchProof:Step3a} are described above in~\eqref{Eqn:AchProof:Step3aa}. Notes for~\eqref{Eqn:AchProof:Step3aa}:
\begin{enumerate}
\item[a.] expand the event that the \rt{source} $\b{X}$ appears in $\set{C}_X$; 
\item[b.] union bound; 
\item[c.] Bayes' law and $\Pr[(\b{X},\b{Y}_k) \in \set{T}^n_\es(P_{XY_k})]\leq 1$; 
\item[d.] conditioned on $(\b{X},\b{Y}_k)$ typical, $\b{X} = \b{X}(m)$ and $\b{W} = \b{W}(m)$, the error $\b{X} \notin \set{L}_k$ occurs if and only if $(\b{W},\b{U}_k)$ are not jointly typical; 
\item[e.] the source and channel codebooks are independent, and all channel codewords are constructed in the same way; 
\item[f.] Lemma~\ref{Lem:Kramer-1}; and 
\item[g.] bound the codebook cardinality $M$ as in~\eqref{Eqn:M}.
\end{enumerate}
Combining~\eqref{Eqn:AchProof:Step2} and~\eqref{Eqn:AchProof:Step3a} with~\eqref{Eqn:AchProof:Step1} and Lemma~\ref{Lem:Kramer-1}, we have
\setcounter{equation}{28}
\begin{equation}\label{Eqn:AchProof:Step3}
\Pr[\b{X} \notin \set{L}_k] \leq b_3\ 2^{-a_3 n},
\end{equation}
for some finite $a_3,b_3 > 0$. 


\subsection{Error Analysis: \rt{Probability receiver $k$'s list $\set{L}_k$ is too large}} 

Now consider the probability that \emph{the size of list $\set{L}_k$ is too large}. We start with
\begin{align}
\Pr\big[|\set{L}_k| > 2^{nD_k}\big]
\notag
&\leq
\Pr\big[\b{X} \notin \set{C}_X\big]
+ \Pr\big[ (\b{X},\b{Y}_k) \notin \set{T}^n_{\es_1} \big]\\
\notag
&\hspace{10mm}
+ \Pr\big[ (\b{W},\b{U}_k) \notin \set{T}^n_{\ec_1} \big]\\
\label{Eqn:AchProof:LS:Step1}
&\hspace{10mm} 
+ \Pr\big[ |\set{L}_k| > 2^{nD_k} \big| \set{S}_2 \big],
\end{align} 
where 
\begin{equation*}
\set{S}_2 := \big\{ \b{X} \in \set{C}_X\big\}
\cap
\big\{ (\b{X},\b{Y}_k) \in \set{T}^n_{\es_1} \big\} 
\cap \big\{ (\b{W},\b{U}_k) \in \set{T}^n_{\ec_1} \big\},
\end{equation*}  
and $\set{T}^n_{\es_1}(P_{X,Y_k})$ and $\set{T}^n_{\ec_1}(P_{W,U_k})$ have been abbreviated by $\set{T}^n_{\es_1}$ and $\set{T}^n_{\ec_1}$ respectively. Apply Markov's inequality to the rightmost probability in~\eqref{Eqn:AchProof:LS:Step1} to get
\begin{equation}\label{Eqn:AchProof:LS:Step2} 
\Pr\big[ |\set{L}_k| > 2^{nD_k} \big| {\set{S}_2} \big]
\leq 
2^{-nD_k}\ 
\E\big[ |\set{L}_k| \big| {\set{S}_2} \big],
\end{equation}
where the expectation is understood to be 
\begin{equation*}
\E\big[ |\set{L}_k| \big| {\set{S}_2} \big] := \sum_{l} l \cdot \Pr\big[ |\set{L}_k| = l \big| {\set{S}_2}\big].
\end{equation*}
We now expand the above expectation over $\b{X} \in \set{C}_X$ (the $M$ possible encodings of $\b{X}$) to get 
\begin{align}
\notag
&\E\big[ |\set{L}_k| \big| \set{S}_2 \big]\\
\notag
&\!\! = \sum_{m = 1}^M 
\E\Big[ |\set{L}_k| \Big| \set{S}_2 \cap \{\b{X} \neq \b{X}(m'), \forall m' < m \} \cap \{\b{X} = \b{X}(m)\} \Big]\\
\label{Eqn:AchProof:LS:Step3} 
&\hspace{7mm}
\cdot \Pr\Big[ \{\b{X} \neq \b{X}(m'),\forall m' < m \} \cap \{\b{X} = \b{X}(m)\} \Big| \set{S}_2 \Big].
\end{align}
Consider the expectation on the right hand side of~\eqref{Eqn:AchProof:LS:Step3}. Let
\begin{equation*}
{\set{S}_{2,m}} := {\set{S}_2} \cap \{ \b{X} \neq \b{X}(m'),\forall m' < m\} \cap \{\b{X} = \b{X}(m)\}.
\end{equation*}
We have
\begin{align}
\notag
&\E\big[ |\set{L}_k| \big| {\set{S}_{2,m}} \big]\\
\notag
& = 
\E \Bigg[\!
\sum_{\tilde{m} = 1}^M\! 
\indicator{(\b{X}(\tilde{m}),\b{Y}_k) \in \set{T}^n_\es}
\indicator{(\b{W}(\tilde{m}),\b{U}_k) \in \set{T}^n_\ec}
\Bigg| {\set{S}_{2,m}} \Bigg]\\
\notag
& =
\sum_{\tilde{m} = 1}^M
 \Pr \Big[ 
\{(\b{X}(\tilde{m}),\b{Y}_k) \in \set{T}^n_\es\}
\cap \{(\b{W}(\tilde{m}),\b{U}_k) \in \set{T}^n_\ec\}
\Big|
{\set{S}_{2,m}} \Big]\\
\notag
& =
\sum_{\tilde{m} = 1}^M
 \Pr \big[ 
(\b{X}(\tilde{m}),\b{Y}_k) \in \set{T}^n_\es
\big|
{\set{S}_{2,m}} \big]\\
\label{Eqn:AchProof:LS:Step5} 
&\hspace{3mm}
\cdot \Pr \big[ 
(\b{W}(\tilde{m}),\b{U}_k) \in \set{T}^n_\ec
\big|
{\set{S}_{2,m}}
\cap \{(\b{X}(\tilde{m}),\b{Y}_k) \in \set{T}^n_\es\}
\big],
\end{align}
where $\set{T}^n_\es(P_{X,Y_k})$ and $\set{T}^n_\ec(P_{W,U_k})$ have been abbreviated by $\set{T}^n_\es$ and $\set{T}^n_\ec$ respectively. 

The event ${\set{S}_{2,m}}$ implies that the \rt{source} $\b{X}$ is equal to the $m$-th source codeword $\b{X}(m)$ and $\b{W} = \b{W}(m)$ is sent over the channel. We now bound the two probabilities on the right hand side of~\eqref{Eqn:AchProof:LS:Step5} separately for each of the three cases $1 \leq \tilde{m} < m$, $\tilde{m} = m$ and $m < \tilde{m} \leq M$. 

\emph{Case 1 $(1 \leq \tilde{m} < m)$:} The first probability in~\eqref{Eqn:AchProof:LS:Step5} is bounded by
\begin{align}
\notag
&\Pr \big[ 
(\b{X}(\tilde{m}),\b{Y}_k) \in \set{T}^n_\es
\big|
{\set{S}_{2,m}} \big]\\
\notag
&\hspace{5mm}
\step{a}{=} 
\Pr\big[
(\b{X}(\tilde{m}),\b{Y}_k) \in \set{T}^n_\es
\big| 
\{(\b{X},\b{Y}_k) \in \set{T}^n_{\es_1}\} 
\cap 
\{ \b{X} \neq \b{X}(\tilde{m})\} \big] \\
\notag
&\hspace{5mm}
\step{b}{\leq}
\frac{\Pr\big[(\b{X}(\tilde{m}),\b{Y}_k) \in \set{T}^n_\es(P_{X,Y_k}) \big| \b{Y}_k \in \set{T}^n_{\es_1}(P_{Y_k})\big]}
{\Pr\big[ \b{X} \neq \b{X}(\tilde{m}) \big| \b{X} \in \set{T}^n_{\es_1}(P_X)\big]}\\
\label{Eqn:AchProof:LS:Step6} 
&\hspace{5mm}
\step{c}{\leq} 
\alpha_n\ 
2^{-n(I(X;Y_k)-2\es H(X))}.
\end{align}
Notes: 
\begin{enumerate}
\item[a.] codewords and codebook are generated independently; 
\item[b.] Bayes' law and the trivial bound $\Pr\big[\b{X} \neq \b{X}(\tilde{m}) \big|\{(\b{X}(\tilde{m}),$ $\b{Y}_k) \in \set{T}^n_\es(P_{X,Y_k})\} \cap \{(\b{X},\b{Y}_k) \in \set{T}^n_{\es_1}\}\big]\leq 1$; and 
\item[c.] apply Lemmas~\ref{Lem:Kramer-1} and~\ref{Lem:Kramer-2} respectively to the denominator and numerator in step (b) and set
\begin{equation}\label{Eqn:alpha}
\alpha_n := \frac{\exp(2^{nH(X)(1-\es_1)})}
{\exp(2^{nH(X)(1-\es_1)})-1}.
\end{equation}
\end{enumerate}
Similarly, by Lemma~\ref{Lem:Kramer-2}, the rightmost probability in~\eqref{Eqn:AchProof:LS:Step5} is bounded by
\begin{multline}\label{Eqn:AchProof:LS:Step7} 
\Pr \big[ 
(\b{W}(\tilde{m}),\b{U}_k) \in \set{T}^n_\ec
\big|
{\set{S}_{2,m}}
\cap \{(\b{X}(\tilde{m}),\b{Y}_k) \in \set{T}^n_\es\}
\big]\\
\leq
2^{-n(I(W;U_k) - 2 \ec H(W))}.
\end{multline}

\emph{Case 2 $(\tilde{m} = m)$:} Bound both probabilities in~\eqref{Eqn:AchProof:LS:Step5} by one. 

\medskip

\emph{Case 3 $(m < \tilde{m} \leq M)$:} {Conditioned on ${\set{S}_{2,m}}$, the encoder has only considered the codewords $\b{X}(1),\ldots, \b{X}(m)$. Thus, even conditional on ${\set{S}_{2,m}}$, the codewords  thereafter $\b{X}(m+1),\ldots, \b{X}(M)$ are independent iid $\sim$ $P_X$ sequences. From Lemma~\ref{Lem:Kramer-2},} 
\begin{equation}\label{Eqn:AchProof:LS:Step8} 
\Pr \big[ 
(\b{X}(\tilde{m}),\b{Y}_k) \in \set{T}^n_\es
\big|
{\set{S}_{2,m}} \big]
\leq 
2^{-n(I(X;Y_k) - 2 \es H(X))},
\end{equation}
Similarly, 
\begin{multline}
\label{Eqn:AchProof:LS:Step9} 
\Pr \big[ 
(\b{W}(\tilde{m}),\b{U}_k) \in \set{T}^n_\ec
\big|
{\set{S}_{2,m}} \cap \{(\b{X}(\tilde{m}),\b{Y}_k)
 \in \set{T}^n_\es\}
\big]\\ 
\leq
2^{-n(I(W;U_k) - 2 \ec H(W))}.
\end{multline}

Collectively, \eqref{Eqn:M} and \eqref{Eqn:AchProof:LS:Step5} to~\eqref{Eqn:AchProof:LS:Step9} imply  
\begin{multline}
\label{Eqn:AchProof:LS:Step4}  
\E\big[ |\set{L}_k| \big| {\set{S}_{2,m}} \big]\\
\leq 1 + 
\alpha_n
2^{n(H(X|Y_k) -  I(W;U_k))} 2^{n(3\es H(X) + 2\ec H(W))}.
\end{multline}
{Combine~\eqref{Eqn:AchProof:LS:Step1}, \eqref{Eqn:AchProof:LS:Step2}
and \eqref{Eqn:AchProof:LS:Step4} to get}
\begin{align}\label{Eqn:AchProof:LS:Step8}
\notag
&\!\!\! \Pr\big[ |\set{L}_k| > 2^{nD_k} \big]\\
\notag
&\leq 
\Pr[\b{X} \notin \set{C}_X] 
+ \Pr[(\b{X},\b{Y}_k) \notin \set{T}^n_\es] + \Pr[(\b{W},\b{U}_k) \notin \set{T}^n_\ec]\\
\notag
&\hspace{5mm}
+ \alpha_n 2^{-n ( D_k - H(X|Y_k) + \rt{I(W;U_k)} )} 
2^{n (3 \es H(X) + 2 \ec H(W))}\\
&\hspace{5mm}
+ 2^{-nD_k}.
\end{align}
{Lemma~\ref{Lem:Kramer-1} and~\eqref{Eqn:AchProof:Step1} imply}
\begin{equation}\label{Eqn:AchProof:LS:FinalBound}
\Pr\big[ |\set{L}_k| > 2^{nD_k} \big] \leq b_4\ 2^{-a_4 n},
\end{equation}
for some finite $a_4,b_4 > 0$ whenever 
\begin{equation*}
D_k > \max\{H(X|Y_k) - I(W;U_k),0\} + 3 \es H(X) + 2 \ec H(W)
\end{equation*}
and $\es,\es_1,\ec$ and $\ec_1$ satisfy~\eqref{Eqn:epsilons}. The result follows because $\es$ and $\ec$ can be chosen arbitrarily small and $H(X)$ and $H(W)$ are finite.
\hfill $\blacksquare$


\section{Proof of Theorem~\ref{Thm:M1} --- Converse}\label{Sec:Proof:Thm:M1:Con}

\rt{Fix $\epsilon > 0$.} Consider any $(\rt{\ns,\nc},R_1,R_2,\ldots,R_K)$-code with $\Pr[\hat{\b{X}}_k \neq \b{X}] \leq \epsilon$ for all $k$. Recall the timeshared pmf $P_{\tilde{W}}$ on $\set{W}$, defined in~\eqref{Eqn:TimeshareCon}. Let 
\begin{equation*}
(\tilde{W},\tilde{U}_1,\tilde{U}_2,\ldots,\tilde{U}_K) \sim P_{\tilde{W}}(\cdot)\ T(\cdot|\cdot)
\end{equation*}
and $\tilde{V}_k =$ $\phi_k(\tilde{W})$. Mirroring the steps of Section~\ref{Sec:Proof:Lem:SW-List:Con}: 
\begin{align}
\notag
 \rt{\nc} I(\tilde{W};\tilde{U}_k,\tilde{V}_k) 
&\geq I(\b{W};\b{U}_k,\b{V}_k)
\notag
\step{a}{\geq} I(\b{W};\b{U}_k,M_k)\\
\notag
&\step{b}{\geq} I(\b{X};\b{U}_k,M_k|\b{Y}_k)\\
\label{Eqn:Thm:M1:Con1}
&= \rt{\ns} H(X|Y_k) - H(\b{X}|\b{Y}_k,\b{U}_k,M_k),
\end{align}
where (a) and (b) use $M_k \leftrightarrow (\b{U}_k,\b{V}_k) \leftrightarrow \b{W}$ and $(\b{X},\b{Y}_k) \leftrightarrow \b{W} \leftrightarrow (\b{U}_k,M_k)$. Similarly, 
\begin{align}
\notag
 \rt{\nc} I(\tilde{W};\tilde{U}_k) &+ \rt{\ns}R_k\\
\notag
&\geq I(\b{W};\b{U}_k) + H(M_k|\b{U}_k)\\
\notag
&\geq I(\b{W};\b{U}_k,M_k)\\
\label{Eqn:Thm:M1:Con2}
&\geq \rt{\ns} H(X|Y_k) - H(\b{X}|\b{U}_k,\b{Y}_k,M_k).
\end{align} 

After applying Fano's inequality~\cite[Thm.~2.10.1]{Cover-2006-B} to $H(\b{X}|\b{U}_k,\b{Y}_k,M_k)$ in~\eqref{Eqn:Thm:M1:Con1} and~\eqref{Eqn:Thm:M1:Con2}, the converse follows in the same way as the closing of Section~\ref{Sec:Proof:Lem:SW-List:Con}. \hfill $\blacksquare$


\begin{figure*}[b!]
\setcounter{equation}{56}
\input{./DoubleColumnEqns/StepsFor47}
\end{figure*}

\begin{figure*}
\input{./DoubleColumnEqns/StepsFor47part2}
\hrule
\end{figure*}
\setcounter{equation}{43}

\section{Proof of Theorem~\ref{Thm:M1} {---} Achievability}\label{Sec:Proof:Thm:M1:Ach}

\rt{We now present an achievability proof for the bandwidth matched case, where $\kappa = 1$ and $\ns = \nc = n$. The mismatched bandwidth case follows by similar arguments.} Our approach to the proof combines the list decoder of Section~\ref{Sec:Proof:Lem:SW-List:Ach} with hash-and-forward coding at the helpers.

\subsection{Code Construction}

Fix a pmf $P_W$ on $\set{W}$ and let us assume that for all $k$ 
\begin{equation}\label{Eqn:Thm:M1:Assumption1}
H(X|Y_k) < I(W;U_k) + \min\big\{R_k,I(W;V_k|U_k)\big\}
\end{equation}
and 
\begin{equation}\label{Eqn:Thm:M1:Assumption2}
H(X|Y_k) \geq I(W;U_k).
\end{equation}
The assumption above~\eqref{Eqn:Thm:M1:Assumption1} matches that in Theorem~\ref{Thm:M1}, and~\eqref{Eqn:Thm:M1:Assumption2} ensures that every receiver requires a positive helper rate to reliably decode the source $\b{X}$. At the end of the proof, we will relax~\eqref{Eqn:Thm:M1:Assumption2} to include situations where some receivers don't require a positive helper rate, i.e., $H(X|Y_k) < I(W;U_k)$ for some $k$.

Generate a random list code, as described in Section~\ref{Sec:Proof:Lem:SW-List:Ach}, with the parameters described above, and let $\set{C}_X$ and $\set{C}_W$ denote the source and channel codebooks respectively. Fix $\es, \es_1, \ec$ and $\ec_1$ arbitrarily small, but always satisfying~\eqref{Eqn:epsilons}. For each receiver~$k$, choose any list exponent $D_k$ satisfying  
\begin{multline}\label{Eqn:DkThm2}
H(X|Y_k) - I(W;U_k) < D_k \\
< I(W;V_k|U_k) - 4 \ec H(W),
\end{multline}
and set the helper rate to be 
\begin{equation}\label{Eqn:HelperRate}
R_k = D_k + \eh
\end{equation}
for some arbitrarily small $\eh > 0$. Notice that it is always possible to choose $D_k$ in~\eqref{Eqn:DkThm2} because~\eqref{Eqn:Thm:M1:Assumption1} and~\eqref{Eqn:Thm:M1:Assumption2} imply $H(X|Y_k) < I(W;U_k,V_k)$ and $I(W;V_k|U_k) > 0$; we can choose $\ec$ arbitrarily small; and $H(W)$ is finite.

We construct a random codebook for helper BS($k$): The codebook is generated by applying the map $\phi_k$ symbol-by-symbol to each codeword $\b{W}(m) \in \set{C}_W$; that is,
\begin{equation*}
\set{C}_{V_k} := \bigcup_{m = 1}^M \big\{\phi_k(\b{W}(m))\big\},
\end{equation*}
where 
\begin{equation*}
\phi_k(\b{W}(m)) = \big(\phi_k(W_1(m)),\phi_k(W_2(m)),\ldots,\phi_k(W_n(m))\big)
\end{equation*}
is a slight abuse of notation.

Uniformly at random place each codeword in $\set{C}_{V_k}$ into one of $\lceil 2^{n R_k} \rceil$ bins. Uniquely label each bin with an index from the set $\{1,2,\ldots,\lceil 2^{nR_k} \rceil\}$, and let $f_k(\b{v})$ denote the bin index of the codeword $\b{v} \in \set{C}_{V_k}$. Denote the set of all codewords in the $b$-th bin by $\set{B}_k(b) := \big\{ \b{v} \in \set{C}_{V_k} : f_k(\b{v}) = b \big\}$
for $b \in \{1,2,\ldots,\lceil 2^{nR_k} \rceil\}$.


\subsection{Encoding and Decoding}

The list encoder and decoders operate as before, see Sections~\ref{Sec:ListEncoder} and~\ref{Sec:ListDecoder}. Helper BS($k$) looks for $\b{V}_k = \phi_k(\b{W})$ in $\set{C}_{V_k}$ and, if successful, sends the bin index $B = f_k(\b{V}_k)$ to receiver $k$. If unsuccessful, the helper sends an index with an independent and uniform distribution. 

The list decoder at receiver $k$ outputs $\set{L}_k$, see~\eqref{Eqn:List}, from which the receiver computes a new list of $V_k$-codewords:
\begin{equation*}
\set{L}_k^* := \Big\{ 
\b{v} \in \set{C}_{V_k} : 
\exists\ \b{X}(m) \in \set{L}_k \text{ with } \b{v} = \phi_k\big(\b{W}(m)\big) \Big\}.
\end{equation*}
If there is a unique codeword $\b{v}'$ in the intersection {of} the list $ \set{L}^*_k$ and the bin $\set{B}_k(B)$, then receiver $k$ sets $\hat{\b{V}}_k := \b{v}'$. Otherwise, receiver $k$ generates $\hat{\b{V}}_k$ iid $\sim P_{V_k}$. 

Finally, receiver $k$ looks for a unique source codeword $\b{X}(m') \in \set{L}_k$ such that $(\b{W}(m'),\b{U}_k,\hat{\b{V}}_k) \in \set{T}_\es(P_{W,U_k,V_k})$.
If successful, the receiver outputs $\hat{\b{X}}_k := \b{X}(m')$; otherwise, it selects a codeword $\b{X}(m)$ uniformly at random from $\set{C}_X$.


\subsection{Error Analysis}

To bound the probability that receiver $k$ decodes in error, $\Pr[\hat{\b{X}}_k \neq \b{X}]$, it is useful to start with
\begin{equation}\label{Eqn:M1:Ach:Step1a}
\Pr[\hat{\b{X}}_k \neq \b{X}] 
\leq 
\Pr[\hat{\b{V}}_k \neq \b{V}_k ] 
+ \Pr[\{\hat{\b{V}}_k = \b{V}_k \} \cap \{\hat{\b{X}}_k \neq \b{X}\} ].
\end{equation}
We may bound the probability that receiver $k$ incorrectly decodes $\b{V}_k$ by conditioning on the list error event $\set{E}$, defined in~\eqref{Eqn:LDError}, and the encoder error $\{\b{X} \notin \set{C}_X\}$ as follows:
\begin{multline}\label{Eqn:M1:Ach:Step1b}
\Pr[\hat{\b{V}}_k \neq \b{V}_k]
\leq 
\Pr[\set{E}]  
+\Pr[\b{X} \notin \set{C}_X]\\
+\Pr\big[\hat{\b{V}}_k \neq \b{V}_k \big| \set{E}^c \cap \{\b{X} \in \set{C}_X\}\big].
\end{multline}
Upper bounds for $\Pr[\set{E}]$ and $\Pr[\b{X} \notin \set{C}_X]$ are given in~\eqref{Eqn:Error:AchProof} and~\eqref{Eqn:AchProof:Step1} respectively. Let us now rewrite the conditional probability in~\eqref{Eqn:M1:Ach:Step1b} using the law of total probability as
\begin{align}
\notag
&\Pr\Big[\hat{\b{V}}_k \neq \b{V}_k\Big|\set{E}^c \cap \{\b{X} \in \set{C}_X\}\Big]\\
\notag
&=
\sum_{b=1}^{\lceil 2^{nR_k}\rceil} \Pr\Big[\{f_k(\b{V}_k)=b\} \cap \{\hat{\b{V}}_k \neq \b{V}_k\}\Big|\set{E}^c \cap \{\b{X} \in \set{C}_X\}\Big]\\
\notag
&= 
\sum_{b=1}^{\lceil 2^{nR_k}\rceil}  \Pr\Big[ f_k(\b{V}_k)=b\Big| \set{E}^c \cap \{\b{X} \in \set{C}_X\} \Big]\\ 
\label{Eqn:M1:Ach:Step2}
&\hspace{10mm}
\cdot \Pr\Big[\hat{\b{V}}_k \neq \b{V}_k \Big| \{f_k(\b{V}_k)=b\}  \cap \set{E}^c \cap \{\b{X} \in \set{C}_X\}\Big].
\end{align}
We now fix a bin $b$ and derive
\begin{align}
\notag
& \Pr\Big[\hat{\b{V}}_k \neq \b{V}_k \Big|  \{f_k(\b{V}_k)=b\}  \cap \set{E}^c \cap \{\b{X} \in \set{C}_X\}\Big]\\
\notag
&\step{a}{=} 
\Pr\Big[ 
\hspace{-2mm}
\bigcup_{\substack{\b{v} \in \set{L}^*_k\\ \b{v} \neq \phi_k(\b{W})}} 
\hspace{-2mm}
\big\{ f_k(\b{v}) = b \big\}
\Big|\{f_k(\b{V}_k)=b\}  \cap \set{E}^c \cap \{\b{X} \in \set{C}_X\}\Big]\\
\notag
&\step{b}{\leq} 
\hspace{-2mm}
\sum_{\substack{\b{v} \in \set{L}_{k}^*\\ \b{v} \neq \phi_k(\b{W})}} 
\hspace{-2mm}
\Pr\Big[ 
f_k(\b{v}) = b
\Big|
 \{f_k(\b{V}_k)=b\}  \cap \set{E}^c \cap \{\b{X} \in \set{C}_X\}
\Big]\\
\notag
&\step{c}{=} 
\sum_{\substack{\b{v} \in \set{L}_{k}^*\\ \b{v} \neq \phi_k(\b{W})}} 
\frac{1}{\lceil 2^{n R_k}\rceil}\\
\notag
&\step{d}{\leq} 
2^{-n (R_k - D_k)}\\
\label{Eqn:M1:Ach:Step4a}
&\step{f}{=} 
2^{-n \eh}.
\end{align}
\newpage
Notes on~\eqref{Eqn:M1:Ach:Step4a}:
\begin{enumerate}
\item[a.] receiver $k$ decodes $\b{V}_k$ in error if and only if there is another $\b{v} \in \set{L}^*_k$ assigned to the same bin as the correct $\b{v}$-codeword; 
\item[b.] the union bound; 
\item[c.] the codewords in $\set{C}_{V_k}$ are thrown uniformly at random into $\lceil 2^{nR_k}\rceil$ bins; 
\item[d.] $|\set{L}_{V_k}| \leq 2^{nD_k}$, since we condition on $\set{E}^c$; and 
\item[f.] substitute the choice of helper rate $R_k$ in~\eqref{Eqn:HelperRate}. 
\end{enumerate}
The right-hand side of \eqref{Eqn:M1:Ach:Step4a} is independent of $b$, so~\eqref{Eqn:M1:Ach:Step2} gives
\begin{equation}\label{Eqn:M1:Ach:Step4a2}
\Pr\big[\hat{\b{V}}_k \neq \b{V}_k\big|\set{E}^c \cap \{\b{X} \in \set{C}_X\}\big]\leq 2^{-n\eh}.
\end{equation}
Combining~\eqref{Eqn:M1:Ach:Step1b}, \eqref{Eqn:Error:AchProof}, \eqref{Eqn:AchProof:Step1} and~\eqref{Eqn:M1:Ach:Step4a2} gives
\begin{equation}\label{Eqn:M1:Ach:Step4b}
\Pr[\hat{\b{V}}_k \neq \b{V}_k] \leq b_5\ 2^{-a_5 n},
\end{equation}
for some finite $a_5,b_5> 0$.

{We now turn to} the rightmost probability in~\eqref{Eqn:M1:Ach:Step1a}. We have
\begin{equation}
\label{Eqn:M1:Ach:Step4}
\Pr[\{\hat{\b{V}}_k = \b{V}_k\} \cap \{ \hat{\b{X}}_k \neq \b{X}\} ]
\leq \Pr[{\set{S}_3^c}] + \Pr[\hat{\b{X}}_{{k}} \neq \b{X} | {\set{S}_3}],
\end{equation} 
where 
\begin{multline*}
{\set{S}_3} := \{\b{X} \in \set{C}_X\} 
\cap \{(\b{X},\b{Y}_k) \in \set{T}_{\es_1}(P_{X,Y_k})\}
\cap \{\hat{\b{V}}_{{k}} = \b{V}_{{k}}\}\\
\cap \{(\b{W},\b{U}_k,\b{V}_k) \in \set{T}_{\ec_1}(P_{W,U_k,V_k})\}
\cap \{|\set{L}_k| \leq 2^{nD_k}\}.
\end{multline*}

An {upper} bound on the probability $\Pr[{\set{S}_3^c}]$ in~\eqref{Eqn:M1:Ach:Step4} follows easily from previous bounds:
\begin{align}
\Pr[{\set{S}_3^c}] 
\notag
&\leq 
\Pr[\b{X} \notin \set{C}_X] 
+ \Pr[(\b{X},\b{Y}_k) \notin \set{T}_{\es_1}]
+ \Pr[\hat{\b{V}}_k \neq \b{V}_k]\\
\notag
&\hspace{10mm} 
+ \Pr[(\b{W},\b{U}_k,\b{V}_k) \notin \set{T}_{\ec_1}] 
+ \Pr[|\set{L}_k| {>} 2^{nD_k}]\\
\label{Eqn:M1:Ach:Step5}
&\step{*}{\leq} 
b_6\ 2^{-a_6 n}, 
\end{align} 
where (*) holds for some finite $a_6, b_6 > 0$ by~\eqref{Eqn:AchProof:Step1},  \eqref{Eqn:AchProof:LS:FinalBound},~\eqref{Eqn:M1:Ach:Step4b}  and Lemma~\ref{Lem:Kramer-1}.

The rightmost probability in~\eqref{Eqn:M1:Ach:Step4} is bounded by
\begin{equation}\label{Eqn:M1:Ach:Step6}
\hspace{-2mm}
\Pr[\hat{\b{X}}_{{k}} \neq \b{X} | {\set{S}_3}] 
\leq
 \gamma_n\
2^{-n(I(W;V_k|U_k)-D_k + 4 \ec H(W))},
\end{equation}
where
\begin{equation*}
\gamma_n := 
\frac{\exp\Big(2n(1-\ec_1)\frac{(\ec-\ec_1)^2}{1+\ec_1}\mu^2_{W,U_k}\Big)}
{\exp\Big(2n(1-\ec_1)\frac{(\ec-\ec_1)^2}{1+\ec_1}\mu^2_{W,U_k}\Big) - 2|\set{W}||\set{U}_k|}.
\end{equation*}
The steps leading to~\eqref{Eqn:M1:Ach:Step6} are shown above in~\eqref{Eqn:M1:Ach:Step6:Proof}. Notes:
\begin{enumerate}
\item[a.] Write $\Pr[\hat{\b{X}}_k \neq \b{X} | {\set{S}_3}]$ as an expectation over all possible realisations of decoder $k$'s list $\set{L}_k$. We note that $|\set{L}_k | \leq 2^{nD_k}$ with probability one, after conditioning on ${\set{S}_3}$. 

\item[b.] Let
\begin{multline*}
{\set{S}_{3,m}} := 
\{\b{X} \neq \b{X}(m'),\forall m' < m\} 
\cap
\{\b{X} = \b{X}(m)\}\\
\cap
{\set{S}_3}
\cap 
\{\set{L} = l\}.
\end{multline*}
and note that $\b{X} \in \set{C}_X$, after conditioning on ${\set{S}_3}$.

\item[c.] An error may only occur if there is some other index $\tilde{m} \neq m$ such that $\b{X}(\tilde{m}) \in \set{L}_k$ and $(\b{W}(\tilde{m}),\b{U}_k,\hat{\b{V}}_k)$ is jointly typical. Here we note that after conditioning on ${\set{S}_{3,m}}$ the following holds with probability one: the \rt{source} $\b{X}$ equals the $m$-th codeword $\b{X}(m)$ in the source codebook $\set{C}_X$; the $m$-th channel codeword is transmitted $\b{W} = \b{W}(m)$; the \rt{source} and side information $(\b{X},\b{Y}_k)$ are $\es$-jointly typical; $(\b{W},\b{U}_k,\b{V}_k)$ are $\ec$-jointly typical; and $\hat{\b{V}} = \b{V}$. We have also abbreviated $\set{T}_{\es_1}(P_{X,Y_k})$ and $\set{T}_{\ec_1}(P_{W,U_k,V_k})$ as $\set{T}_{\es_1}$ and $\set{T}_{\ec_1}$ respectively.

\item[d.] Apply the union bound. 

\item[e.] The rightmost probability in step (d) is bounded by
\begin{multline*}
\Pr\big[ (\b{W}(\tilde{m}),\b{U}_k,\hat{\b{V}}_k) \in \set{T}_\ec 
\big| 
{\set{S}_{3,m}}\big]\\
\leq \gamma_n\ 
2^{-n(I(W;V_k|U_k) - 4\ec H(W))}.
\end{multline*}
The steps leading to this bound are shown above in~\eqref{Eqn:M1:Ach:bbc}. (See below for detailed notes on each step in~\eqref{Eqn:M1:Ach:bbc}.)

\item[f.] \rt{For each list $l$  that satisfies
$|l| \leq 2^{nD_k}$,
\begin{multline*}
\sum_{\substack{\tilde{m} \in l \\ \tilde{m} \neq m}}
\gamma_n 2^{-n(I(W;V_k|U_k) - 4 \delta H(W))}\\
\leq 
\gamma_n 2^{nD_k} 2^{-n(I(W;V_k|U_k) - 4 \delta H(W))}.
\end{multline*}
Step (f) now follows from~\eqref{Eqn:M1:Ach:bbc:stepf} above.}
\end{enumerate}

\noindent Notes for~\eqref{Eqn:M1:Ach:bbc}:
\begin{enumerate}
\item[e.1.] This step follows from the independence of the source and channel codebooks, the independence of codewords within each codebook, conditioning on $\set{S}_{3,m}$ and $\{\set{L}_k = l\}$ being equivalent to 
\begin{equation*}
(\b{X}(m'),\b{Y}_k) \in \set{T}_{\es}\ \text{and}\ (\b{W}(m'),\b{U}_k) \in \set{T}_\ec,\ \forall m' \in l,
\end{equation*}
and
\begin{equation*}
(\b{X}(m'),\b{Y}_k) {\notin} \set{T}_{\es}\ \text{or}\ (\b{W}(m'),\b{U}_k) {\notin} \set{T}_\ec,\ \forall m' \notin l.
\end{equation*}

\item[e.2.] \rt{Bayes' rule.}

\item[e.3.]  Apply Lemma~\ref{Lem:Kramer-2} to (e.2). 
\end{enumerate}

Thus, 
\setcounter{equation}{59}
\begin{equation}\label{Eqn:M1:Ach:Step7}
\Pr[\hat{\b{X}}_{{k}} \neq \b{X} | {\set{S}_3}] \leq b_7 \ 2^{-a_7n},
\end{equation}
for some $b_7>0$ and $a_7 :=I(W;V_k|U_k)-D_k \rt{-} 4 \ec H(W)$, where $a_7 > 0$ by~\eqref{Eqn:DkThm2}. Whenever~\eqref{Eqn:Thm:M1:Assumption1} and~\eqref{Eqn:Thm:M1:Assumption2} both hold, the achievability achievability of Theorem~\ref{Thm:M1} follows from~\eqref{Eqn:M1:Ach:Step1a} and \eqref{Eqn:M1:Ach:Step4b}, \eqref{Eqn:M1:Ach:Step4}, \eqref{Eqn:M1:Ach:Step5}, and~\eqref{Eqn:M1:Ach:Step7}.

To complete the achievability proof of Theorem~\ref{Thm:M1}, we need only relax the assumption~\eqref{Eqn:Thm:M1:Assumption2} and suppose that $H(X|Y_k) < I(W;U_k)$ for one or more receivers $k$. Such receivers do not require a positive helper rate or list exponent (i.e., we can set $R_k = 0$ and $D_k = 0$), and we can instead impose unique decoding. Indeed, the error analysis in~\cite[Sec.~IV]{Tuncel-Apr-2006-A} shows that the probability of error $\Pr[\hat{\b{X}}_k \neq \b{X}]$ at such receivers decays exponentially in $n$. (The error analysis in~\cite[Sec.~IV]{Tuncel-Apr-2006-A} is valid because we use the same random source and channel codebooks.)
\hfill $\blacksquare$


\section{Proof of Theorem~\ref{Thm:M2} {---} Converse}\label{Sec:Proof:Thm:M2:Con}

\rt{Fix $\epsilon > 0$.} Consider any $(\rt{\ns,\nc},R_1,R_2,\ldots,R_K)$-code with $\Pr[\hat{\b{X}}_k \neq \b{X}] \leq \es$ for all $k$. Following the now familiar path of defining $(\tilde{W},$ $\tilde{U}_1,\tilde{U}_2,\ldots,\tilde{U}_K) \sim P_{\tilde{W}}(\cdot) T(\cdot | \cdot)$, with the timeshared pmf $P_{\tilde{W}}$ given in~\eqref{Eqn:TimeshareCon}, we have
\setcounter{equation}{60}
\begin{align}
\notag
\rt{\nc} I(\tilde{W};\tilde{U}_k) 
\notag
&\step{a}{\geq} 
I(\b{W};\b{U}_k)
\step{b}{=} I(\b{X},\b{Y}_k,M_k,\b{W};\b{U}_k)\\
\notag
&\geq 
H(\b{X}|M_k,\b{Y}_k)
- H(\b{X}|M_k,\b{Y}_k,\b{U}_k) 
\\
\notag
&\step{c}{\geq}
\rt{\sum_{i = 1}^\ns} H(X_i | M_k,\b{Y}_k,X_1^{i-1}) 
- \rt{\ns}\varepsilon(\rt{\ns})\\ 
\notag
&\step{d}{\geq} 
\rt{\sum_{i = 1}^\ns} H(X_i | M_k,\b{Y}_k,X_1^{i-1},V_{k,1}^{i-1}) 
- \rt{\ns}\varepsilon(\rt{\ns}) \\
\notag
&\step{e}{=}
\rt{\sum_{i = 1}^\ns} H(X_i | M_k,\b{Y}_k,V_{k,1}^{i-1}) 
- \rt{\ns}\varepsilon(\rt{\ns}) \\
\label{Eqn:Converse:M2:Step1}
&\step{f}{=}
\rt{\sum_{i = 1}^\ns} H(X_i | A_{k,i}, Y_{k,i}) 
- \rt{\ns}\varepsilon(\rt{\ns}). 
\end{align}
Notes: 
\begin{enumerate}
\item[a.] Jensen's inequality; 
\item[b.] $(\b{X},\b{Y}_k,M_k) \leftrightarrow \b{W} \leftrightarrow \b{U}_k$ forms a Markov chain; 
\item[c.] Fano's inequality, where $\varepsilon(n)$ $\rightarrow 0$, and the shorthand notation $X_1^{i-1} = (X_1,X_2,$ $\ldots,X_{i-1})$; 
\item[d.] conditioning reduces entropy and the notation $V_{k,1}^{i-1} = (V_{k,1},V_{k,2},\ldots,V_{k,i-1})$; 
\item[e.] $X_i \leftrightarrow (M_k,V_{k,1}^{i-1},\b{Y}_k) \leftrightarrow X_1^{i-1}$ forms a Markov chain \rt{(see below for details)}; and 
\item[f.] substitutes $A_{k,i} := (M_k,Y_{k,1}^{i-1},Y_{k,i+1}^n,V_{k,1}^{i-1})$.
\end{enumerate}

\rt{To see that $X_i \leftrightarrow (M_k,V_{k,1}^{i-1},\b{Y}_k) \leftrightarrow X_1^{i-1}$ forms a Markov chain in step (e) above, we first notice that 
\begin{equation}\label{Eqn:Converse:M2e:MC1}
\big( X_i,M_k,Y_{k,i}^n \big) \leftrightarrow V_{k,1}^{i-1} \leftrightarrow \big(X_1^{i-1},Y_1^{i-1}\big)
\end{equation}
forms a Markov chain because the source and side information are memoryless and $M_k$ is a function only of $\b{V}_k$. The chain~\eqref{Eqn:Converse:M2e:MC1} implies $X_i \leftrightarrow \big(M_k,V_{k,1}^{i-1},Y_{k,i}^n \big) \leftrightarrow \big(X_1^{i-1},Y_1^{i-1}\big)$,
which, in turn, implies $X_i \leftrightarrow \big(M_k,V_{k,1}^{i-1},\b{Y}_k \big) \leftrightarrow X_1^{i-1}$.
}

The bound for helper rate $R_k$ follows a similar argument to that of~\cite[Sec.~15.8]{Cover-2006-B}. Specifically, 
\begin{align}
\notag
\rt{\ns} R_k &\geq  H(M_k) 
\geq 
 I(\b{V}_k;M_k|\b{Y}_k)\\ 
\notag
&= \rt{\sum_{i=1}^\ns} 
I(V_{k,i};M_k|\b{Y}_k,V_{k,1}^{i-1})\\
\notag
&\step{a}{=} \rt{\sum_{i=1}^\ns} 
I(V_{k,i};M_k,Y_{k,1}^{i-1},Y_{k,i+1}^n,V_{k,1}^{i-1}|Y_{k,i})\\
\notag
&\step{b}{=}\rt{\sum_{i=1}^\ns} 
I(V_{k,i};A_{k,i}|Y_{k,i}),
\end{align}
where step (a) follows because the source is memoryless and (b) substitutes $A_{k,i}$. 

\rt{
The source and the side information are iid and $M_k$ is only a function of $\b{V}_k$, so 
\begin{equation}\label{Eqn:Converse:M2e:MC2}
(X_i,Y_i) \leftrightarrow V_{k,i} \leftrightarrow (M_k,\b{V}_k, Y_{k,1}^{i-1}, Y_{k,i+1}^{n}).
\end{equation}
The Markov chain~\eqref{Eqn:Converse:M2e:MC2} implies $
(X_i,Y_i) \leftrightarrow V_{k,i}\leftrightarrow A_{k,i}$, and the converse} follows from standard timesharing arguments, e.g.~see~\cite[p.~578]{Cover-2006-B}. 
\hfill $\blacksquare$


\section{Proof of Theorem~\ref{Thm:M2} --- Achievability}\label{Sec:Proof:Thm:M2:Ach}

The proof combes the list decoder of Section~\ref{Sec:Proof:Lem:SW-List:Ach} with a `helper' source code at BS($k$).


\subsection{Code Construction}

Fix a pmf $P_W$ on $\set{W}$ and auxiliary random variables $(A_1,$ $A_2,\ldots,A_K)$ satisfying $A_k \leftrightarrow V_k \leftrightarrow (X,Y_k)$. Let us assume that 
\begin{equation}\label{Eqn:Thm:M2:Assumption1}
H(X|A_k,Y_k) < I(W;U_k)\quad \forall\ k
\end{equation}
and
\begin{equation}\label{Eqn:Thm:M2:Assumption2}
R_k > I(A_k;V_k|Y_k)\quad \forall\ k.
\end{equation}
As in Section~\ref{Sec:Proof:Thm:M1:Ach} (the achievability proof Theorem~\ref{Thm:M1}), let us also assume that~\eqref{Eqn:Thm:M1:Assumption2} holds so that every receiver requires a positive helper rate. 

Fix constants $\es$, $\es_1$, $\ec$ and $\ec_1$ satisfying~\eqref{Eqn:epsilons}, and \maw{choose {$0 < \es_1 < \eh_1 < \eh < \mu_{A_k,X,Y_k}$}.} Generate a random list code, as described in Section~\ref{Sec:Proof:Lem:SW-List:Ach}, with the parameters described above, and let $\set{C}_X$ and $\set{C}_W$ denote the source and channel codebooks respectively. Fix the list exponents to be
\begin{equation*}
D_k = H(X|Y_k) - I(W;U_k) + \rho,\quad \forall\ k,
\end{equation*}
for any
\begin{equation}\label{Eqn:Rho}
\rho > 3\es H(X) + 2\ec H(W).
\end{equation}

Let $P_{A_k}$ denote the marginal distribution of $A_k$. Randomly generate a source codebook for BS($k$), with codewords of length $n$,  by selecting  symbols from $\set{A}_k$ iid $\sim$ $P_{A_k}$:
\begin{equation*}
\set{C}_{A_k}\! :=\! \Big\{ \b{A}_{\maw{k}}(j,j')\! =\! \big(A_{\maw{k},1}(j,j'),A_{\maw{k},2}(j,j'),\ldots,A_{\maw{k},n}(j,j')\big)\Big\}
\end{equation*} 
where we call $j$ the \emph{bin index} and
\begin{equation*}
j = 1,2,\ldots,\lfloor 2^{n R_k}\rfloor\ \text{ and }\ j' = 1,2,\ldots,\lfloor2^{n (I(A_{\maw{k}};Y_k)-\eh_1)}\rfloor.
\end{equation*}
 

\subsection{Encoding and Decoding}

The list encoder and decoders operate as before, see Sections~\ref{Sec:ListEncoder} and~\ref{Sec:ListDecoder}. Helper BS($k$) searches through the $A_k$-codebook $\set{C}_{A_k}$ for a pair {$(J,J')$} such that $\big(\b{A}_{\maw{k}}\maw{(J,J')},\b{V}_k\big) \in \set{T}_{\eh_1}$.
If successful, BS($k$) sends the smallest such bin index {$J$} to receiver $k$. If unsuccessful, the helper sends an index $J$ with an independent and uniform distribution \maw{over all possible bin indices.} 

Receiver $k$ first attempts to decode $\b{A}_{\maw{k}}(J,J')$ by looking for a unique $\hat{J}$ in the $J$-th bin such that $\big(\b{A}_{\maw{k}}(J,\hat{J}),\b{Y}_k\big) \in \set{T}_{\eh_1}$. If successful, \maw{receiver~$k$} sets $\hat{\b{A}}_k = \b{A}_{\maw{k}}(J,\hat{J})$. Otherwise, \maw{it} randomly selects $\maw{\hat{\b{A}}}_k$ iid $\sim P_{A_k}$. 

The list decoder at receiver $k$ outputs $\set{L}_k$, see~\eqref{Eqn:List}. \maw{Receiver~$k$} looks for a unique $\b{X}(m') \in \set{L}_k$ such that $
\big(\b{X}(m'),\b{Y}_k,\hat{\b{A}}_k\big) \in \set{T}_\es$. If successful, \maw{receiver~$k$} outputs $\hat{\b{X}}_k := \b{X}(m')$. Otherwise, \maw{it} randomly generates $\hat{\b{X}}_{\maw{k}}$ using $P_X$. 


\setcounter{equation}{67}
\begin{figure*}[t!]
\input{./DoubleColumnEqns/Eqn-S4}
\hrule
\end{figure*}

\begin{figure*}
\input{./DoubleColumnEqns/Eqn-S4-step-e}
\hrule
\end{figure*}
\setcounter{equation}{66}

\subsection{Error Analysis}

We first bound the probability of error at receiver $k$ by
\begin{equation*}
\maw{\Pr[\hat{\b{X}}_k \neq \b{X}] \leq 
\Pr[{\set{S}_4^c}] + \Pr[ \hat{\b{X}}_k \neq \b{X} | {\set{S}_4}],}
\end{equation*}
where 
\begin{multline*}
{\set{S}_4} := 
\{\b{X} \in \set{C}_X \}
\cap \{\hat{\b{A}}_k = \b{A}_k\} 
\cap \{|\set{L}_k| \leq 2^{nD_k}\}\\
\cap \{ (\b{X},\b{Y}_k,\b{A}_k) \in \set{T}_{\es_1}\} 
\cap \{(\b{W},\b{U}_k) \in \set{T}_{\ec_1}\},
\end{multline*}
and we have abbreviated the typical sets $\set{T}_{\es_1}(P_{X,Y_k,A_k})$ and $\set{T}_{\ec_1}(P_{W,U_k})$ as $\set{T}_{\es_1}$ and $\set{T}_{\ec_1}$, respectively.

{We have 
\begin{equation}\label{Eqn:BoundforS4c}
\Pr\big[\set{S}_4^c\big] \leq b_7\ 2^{-a_7 n},
\end{equation}
for some finite $a_7,b_7 > 0$. To see~\eqref{Eqn:BoundforS4c}, apply the union bound to $\Pr\big[\set{S}_4^c\big]$; use~\eqref{Eqn:AchProof:Step1a} to bound $\Pr[\b{X} \notin \set{C}_X]$; use~\eqref{Eqn:AchProof:LS:Step8}, ~\eqref{Eqn:AchProof:LS:FinalBound} and~\eqref{Eqn:Rho} to bound $\Pr[|\set{L}_k| > 2^{nD_k}]$; and use  Lemmas~\ref{Lem:Kramer-1} and~\ref{Lem:Kramer-2} to bound $\Pr[(\b{W},\b{U}_k) \notin \set{T}^n_{\ec_1}]$. The final two probabilities, $\Pr[\hat{\b{A}}_k \neq \b{A}]$ and $\Pr[(\b{X},\b{Y}_k,\b{A}_k) \notin \set{T}^n_{\es_1}]$, also tend to zero exponentially in $n$ by Lemmas~\ref{Lem:Kramer-1} and~\ref{Lem:Kramer-2}; see, for example, Kramer's achievabiltiy proof of the Wyner-Ziv theorem \cite[Sec.~5.3]{Kramer-2008-A}. Finally, the conditional probability $\Pr[\hat{\b{X}}_k \neq \b{X}|\set{S}_4]$ tends to zero exponentially in $n$, as shown below in~\eqref{Eqn:Thm2Proof:Stepb1}.}

Notes for~\eqref{Eqn:Thm2Proof:Stepb1}:
\begin{enumerate}
\item[a.] Write $\Pr[\hat{\b{X}}_k \neq \b{X} | \set{S}_4]$ as an expectation over all possible realisations of decoder $k$'s list $\set{L}_k$. Here we note that $|\set{L}_k| \leq 2^{nD_k}$ with probability one, after conditioning on $\set{S}_4$. 

\item[b.] Write the second conditional probability in step (a) as an expectation over all possible encodings of $\b{X}$. Here we note that $\b{X} \in \set{C}_X$ with probability one, after conditioning on $\set{S}_4$. 

\item[c.] In the rightmost conditional probability in step (b), the error event $\{\hat{\b{X}}_k \neq \b{X}\}$ is equivalent to the following: There exists an index $\tilde{m}$ in decoder $k$'s list $\set{L}_k$, which is different to the correct index $m$ \maw{and} such that $(\b{X}(\tilde{m}),\b{Y}_k,\hat{\b{A}}_k)$ is jointly typical. Here we note that the correct index $m$ is in decoder $k$'s list with probability one, after conditioning \maw{on} $\set{S}_4$. We have also slightly abused notation and written the union over all indices $m' \in \set{L}_k$, but $\set{L}_k$ is a list of source codewords, \maw{see}~\eqref{Eqn:List}. 

\item[d.] Apply the union bound to (c). 

\item[e.] If the index $\tilde{m}$ is smaller than $m$, $\tilde{m} < m$, then the rightmost conditional probability in step (d) is bounded from above by~\eqref{Eqn:Thm2Proof:Stepb1.e}, which is given below. It can also be shown that~\eqref{Eqn:Thm2Proof:Stepb1.e} holds for indices $\tilde{m} > m$. To see this note that $\gamma_n > 1$ and the righthand side of step (e.1) in~\eqref{Eqn:Thm2Proof:Stepb1.e} simplifies to 
\begin{multline*}
\Pr\big[  
(\b{X}(\tilde{m}),\b{Y}_k,\b{A}_k) \in \set{T}_\es 
\big|
\{(\b{X}(\tilde{m}),\b{Y}_k) \in \set{T}_\es\}\\ 
\cap \{(\b{X},\b{Y}_k,\b{A}_k) \in \set{T}_{\es_1}\}
\big].
\end{multline*}
The bound then follows from Lemma~\ref{Lem:Kramer-2}.

\item[f.] Substitute $D_k = H(X|Y_k) - I(W;U_k) + \rho$. 
\end{enumerate}

\noindent Notes for~\eqref{Eqn:Thm2Proof:Stepb1.e}:
\begin{enumerate}
\item[e.1.] The first step is a consequence of the independence of the source and channel codebooks, the independence of codewords within each codebook, and $\{\set{L}_k = l\}$ is equivalent to 
\begin{equation*}
(\b{X}(m'),\b{Y}_k) \in \set{T}_\es \text{ and } (\b{W}(m'),\b{U}_k) \in \set{T}_\ec,\quad \forall m' \in l,
\end{equation*} 
and
\begin{equation*}
(\b{X}(m'),\b{Y}_k) \notin \set{T}_\es \text{ or } (\b{W}(m'),\b{U}_k) \notin \set{T}_\ec,\quad \forall m' \notin l.
\end{equation*} 

\item[e.2.] Apply Bayes' law twice and use the upper bound
\begin{multline*}
\Pr\big[\{ \b{X} \neq \b{X}(\tilde{m})\}
\cap \{(\b{X}(\tilde{m}),\b{Y}_k) \in \set{T}_\es\} \big|
\{(\b{X}(\tilde{m}),\\ \b{Y}_k,\b{A}_k) \in \set{T}_\es \}
\cap \{(\b{X},\b{Y}_k,\b{A}_k) \in \set{T}_{\es_1}\}\big] \leq 1
\end{multline*}

\item[e.3.] Use Lemmas~\ref{Lem:Kramer-1} and~\ref{Lem:Kramer-2} to lower bound
\begin{multline*}
1 - 
\Pr\big[\b{X} = \b{X}(\tilde{m}) 
\big| \{(\b{X}(\tilde{m}),\b{Y}_k) \in \set{T}_\es\} \\
\cap \{(\b{X},\b{Y}_k,\b{A}_k) \in \set{T}_{\es_1} \}
\big].
\end{multline*}
Use Lemma~\ref{Lem:Kramer-2} to bound the numerator and denominator of the rightmost term in step (e.2).  

\item[e.4.] Set 
\begin{equation*}
\gamma_n := 
\frac{1}
{
1 - \frac{1}{1 - \zeta_n} 2^{-n (H(X|Y_k) - 3\es H(X))},
}
\end{equation*} 
where
\begin{equation*}
\zeta_n := 2|\set{X}||\set{Y}_k| \exp\Big( -2n(1-\es_1) \frac{(\es-\es_1)^2}{1 +\es_1} \mu^2_{X,Y_k} \Big)
\end{equation*}
We notice that $\gamma_n \rightarrow 1$ from above whenever $3 \es H(X) < H(X|Y_k)$ and $\es > \es_1$.
\end{enumerate}
The proof now follows from~\eqref{Eqn:Thm2Proof:Stepb1.e}, because $H(X|A_k,Y_k) < I(W;U_k)$, we can choose $\es$ and $\rho$ arbitrarily small, and $H(X)$ is finite.\ \hfill $\blacksquare$ 


\section*{Acknowledgements}
The authors would like to thank Gerhard Kramer, the associate editor and the anonymous reviewers for their thoughtful comments on the paper. 


\end{document}